\documentclass[12pt]{article}
\usepackage{amsmath}
\usepackage{amsfonts}
\usepackage{epsfig}
\usepackage{graphicx}
\renewcommand{\Ref}[1]{(\ref{#1})}
\renewcommand{\phi}{\varphi}
\newcommand{\tphi}{\tilde\phi}
\newcommand{\<}{\langle}
\renewcommand{\>}{\rangle}

\newcommand{\abar}{\bar\alpha_s}
\newcommand{\tdm}[1]{\mbox{\boldmath $#1$}}
\newcommand{\pd}{\partial}
\newcommand{\vx}{\tdm{x}}
\newcommand{\vy}{\tdm{y}}
\newcommand{\vz}{\tdm{z}}
\newcommand{\vk}{\tdm{k}}
\newcommand{\vr}{\tdm{r}}
\newcommand{\lra}{\leftrightarrow}

\def\beq{\begin{equation}}
\def\eeq{\end{equation}}

\title{\bf Local one-dimensional reggeon model
of the interaction of pomerons and odderons.}
\author{M.A. Braun, E.M. Kuzminskii, M.I. Vyazovsky\\
Dept. of High Energy physics,
Saint-Petersburg State University,\\
198504 S.Petersburg, Russia}

\begin{document}

\maketitle
{\bf Abstract}

\noindent
We propose the one-dimensional reggeon theory describing local
pomerons and odderons. It generalizes the well-known one-dimensional
theory of pomerons (the Gribov model) and includes only triple
interaction vertices. The proposed theory is studied by numerical
methods: the one-particle pomeron and odderon propagators and
the $pA$ amplitude are found as functions of rapidity by integrating
the evolution equation.

%%1
\section{Introduction}

\noindent
%%%%%%%%%%%%%%%%%%%%%%%%%%%%%%%%%%%%%%%%%%%%%%%%%%%%%%%%%%%%%%%%%%%%
In the framework of the Quantum Chromodynamics in the kinematic region
where energy is much greater than the transferred momenta (''the Regge
kinematics'') the strong interactions can be described by the exchange of
pomerons, which can be interpreted  as bound states of pairs of the
so-called reggeized gluons. In the quasiclassical approximation (which
neglects pomeron loops) and in the approximation of a large number of
colors, for the scattering of a  small projectile off a large target
(''dilute-dense scattering'') it leads to the well-known
Balitsky-Kovchegov (BK) equation widely used for the description of DIS
and particle-nucleus ($pA$) scattering. The BK equation corresponds
to summing fan diagrams going from the projectile to the nuclear target
with the propagator given by the well-known BFKL equation and the triple
pomeron vertices responsible for the splitting of a pomeron in two.
Going beyond the quasiclassical approximation and taking account
of loops presents a hardly surmountable problem, which has been not
solved until now.

In connection with this difficulty much attention was given to the
previous attempts to study the strong interactions before the era of the
QCD and using the old reggeon theory \cite{gribov,migdal,migdal1}
introduced by V.N.Gribov and based on the phenomenological local pomeron
and its interaction vertices. In fact the $pA$ interaction in this
framework was considered in \cite{schwimmer} where the sum of all
fan diagrams similar to the BK equation in the QCD was found.
Unlike the QCD, in the local pomeron model both the pomeron intercept
and coupling constant for the triple pomeron vertex are taken
as phenomenological parameters adjusted to the experimental data.
Needless to say the local pomeron is much poorer in his physical
content as compared to its QCD counterpart, which makes it unfit
to describe processes with hard momentum transfer like DIS.
However, the local pomeron theory is much simpler than the QCD one
and admits various methods which make it possible to go beyond
the quasiclassical approximation. In particular renormalization group
methods have been widely applied to establish some basic properties
of the Gribov model. In  \cite{migdal1,ref1,ref2,ref3}
a reggeon field theory with zero renormalized reggeon mass has been investigated
by means of the Callan-Symanzik equation and the $\epsilon$-expansion was studied
in the vicinity ($D=4-\epsilon$) of the critical dimension of transverse space.
The key result was the existence of an infrared fixed point which leads to scaling
laws for the pomeron Green function. More recently a general reggeon field theory
as investigated in the framework of the functional renormalization group formalism
with the aim of finding fixed points for the coupling constants flows, which may
shed light on comparing it with the QCD and relevance for the description
of the region of low momenta ~\cite{BCV1}.

Still in the realistic three-dimensional world even the local pomeron
model does not allow to find the full quantum-mechanical solution of the
problem with the contribution of pomeron loops fully taken into account.
In view of this trouble a still simpler model (''toy''  model)
was considered and studied in some detail. The local pomeron model was
taken in the one-dimensional world, that is depending only on rapidity $y$
\cite{amati1,amati2,jengo,amati3,rossi}.
Such a model essentially was equivalent to the standard Quantum
Mechanics and studied by methods of the quantum mechanics. Later also
numerical methods were used both for the evaluation of the eigenvalues
of the quantum Hamiltonian \cite{bondarenko,braun3} and for integrating
the partial differential equation which governs the evolution of this
system in rapidity \cite{braun1,braun2}. The important message which
follows from these studies is that the quantum effects, that is the
loops, change cardinally the high-energy behaviour of the amplitudes
and so the neglect of them is at most a very crude approximation.
%%%%%%%%%%%%%%%%%%%%%%%%%%%%%%%%%%%%%%%%%%%%%%%%%%%%%%%%%%%%%%%%%%%%

It is remarkable that in the QCD, apart from the pomeron with the
positive $C$-parity and signature, a compound state of three reggeized gluons with
the negative $C$-parity and signature, the odderon, appears. Its possible experimental
manifestations has not been found with certainty up to now, which may be
explained both by its behaviour with energy and its small coupling
to participant hadrons. On the theoretical level two species of the odderon
were found, the Bartels-Lipatov-Vacca (BLV) odderon \cite{blv} with the
intercept exactly equal to unity, in which the three reggeized gluons
are pairwise located at the same spatial point, and the more
complicated Janik-Wosiek odderon \cite{jw1,jw2} with all three
reggeized gluons at different points, the intercept somewhat below unity
and so probably subdominant at high energies. It was noted
in \cite{hiim} that the BLV odderon is in a certain sense an imaginary
part of the full $S$-matrix with both $C=\pm 1$ exchanges whose real part
is the pomeron. Having this in mind, in analogy with the BK equation
a system of equations summing fan diagrams made of both pomerons $P$
and odderons $O$ has been derived \cite{ksw, hiim}. It takes into account
transitions $P\lra PP$, $P\lra OO$ and $O\lra PO$ with the coupling
constants related as 1:-1:2, respectively. This system is more
complicated than the BK equation but can be solved numerically
as the latter \cite{hiim,motyka,braun4}. However, going beyond
the quasiclassical approximation remains unrealizable.
In the framework of the local reggeon model the odderon has been introduced
in ~\cite{BCV2} again  within the functional renormalization group formalism
and the same goal of finding possible fixed points for the coupling constants
flows. However, even forgetting strong limitations on the number of terms
in the effective action, the found fixed points tell nothing about the actual
values for amplitudes with given phenomenological parameters.

In view of this and learning the lesson from the studies of the theory
with only pomerons, in this paper we propose a generalization of the
one-dimensional ''toy'' model to include interactions of both the
pomerons and odderons.
Note that earlier results found  in the functional renormalization group formalism
in ~\cite{BCV2} do not help much to construct such a model due to the essential
use of infrared regulators inherent in this formalism.
The Lagrangian function of the one-dimensional
model with only the pomerons is
\begin{equation}
{L}=\Phi^{+} \partial_y \Phi - \mu_P \Phi^{+} \Phi
+ i \lambda \Phi^{+}  ( \Phi + \Phi^{+} ) \Phi,
\label{e11}
\end{equation}
where $\Phi,\Phi^{+}$ are the complex pomeron field and its conjugate,
the mass parameter $\mu_P=\alpha(0)-1$ is defined by the intercept
of the pomeron Regge trajectory and $\lambda$ is the effective coupling
constant.

To generalize \Ref{e11} we are guided by the properties
of the transition vertices $P\lra PP$, $P\lra OO$ and $O\lra PO$ found
both in the analysis of the loop diagrams in the QCD ~\cite{BR} and in the quasiclassical
equations~\cite{ksw,hiim}. As derived in ~\cite{BR} the one loop corrections should be
negative  when a pomeron is present in the loop and positive for the odderon loop
in the pomeron propagator. From ~\cite{hiim} we deduce that the intercept of the odderon
is unity and relation of the coupling constants $P\to PP$, $P\to OO$ and $O\to PO$
should be as in the quasiclassical equations. We have also to take into account
the negative signature of the odderon, which makes its contribution
to the amplitude real in contrast to the positive imaginary one for
the pomeron. As will be discussed in the next Section, these properties
dictate the form of the Lagrangian in our new local reggeon model as
$$
{L}= \Phi^{+} \partial_y \Phi - \mu_P \Phi^{+} \Phi
+ i(\Psi^{+} \partial_y \Psi - \mu_O \Psi^{+} \Psi)
$$
\begin{equation}
+ i \lambda
\left( \Phi^{+}\Phi^{+}\Phi + \Phi^{+}\Phi\Phi
+2i \Psi^{+} \Psi \Phi +2i \Phi^{+} \Psi^{+} \Psi
- \Phi^{+} \Psi \Psi - \Psi^{+} \Psi^{+} \Phi \right) .
\label{e12a}
\end{equation}
Here $\Phi$ is the complex pomeron field, $\Psi$ is the complex odderon
field, $\Phi^{+}$ and $\Psi^{+}$ are their conjugates, $\mu_P$ and $\mu_O$
are the pomeron and odderon mass parameters equal to the intercepts minus 1,
respectively, the effective coupling constant $\lambda$ is the same as
in the pure pomeron theory \Ref{e11}.
To take the odderon signature into account its free Lagrangian is multiplied
by $i$, which multiplies its propagator by $-i$. Apart from this we have taken
real the coupling constants for $\Psi^{+} \Psi \Phi$ and its conjugate
in contrast with all the rest coupling constants kept positive imaginary.
This is necessary to conform to the QCD loop contributions and quasiclassical
equations \cite{ksw, hiim, BR}.
This theory is formally different with the one introduced earlier in ~\cite{BCV2},
however, we shall see that in fact they are fully equivalent after the appropriate
transformation of field variables.

This theory describes both creation
and absorption of pomerons and odderons and contains interactions only with
even powers of the odderon field, hereby conserving the $C$-symmetry. It also
has the relation between the coupling constants of different transitions
in correspondence with the QCD. In our work we will start with the
arbitrary $\mu_O$, but in the numerical studies we adopt the zero mass
parameter for the odderon in accordance with property of leading BLV odderon.

In the next Section from the Lagrangian function \Ref{e12a} we construct
the quantum Hamiltonian of the model, study the equations of motion
and compare with the quasiclassical (fan) equations in the QCD \cite{ksw,hiim}.
In Section 3 we consider evolution in rapidity of the pomeron and odderon
propagators and $pA$ amplitude in our model and suggest two calculational
schemes for its numerical study: development in powers and point-like
evolution. In Section 4 the numerical results found by these methods
are presented. In Section 5 the effect of introducing of the quartic
interaction in the theory is studied. Section 6 contains some conclusions.

%%2
\section{The Hamiltonian}
\subsection{Quantization}

\noindent
The free part of the Lagrangian \Ref{e12a} has the standard form
and admit the canonical quantization. Treating $y$ as the imaginary
time one finds that the canonical conjugates to the variables $\Phi$
and $\Psi$ are
\begin{equation}
\Phi^{*} = \Phi^{+} , \quad \Psi^{*} = i \Psi^{+} ,
\label{e20}
\end{equation}
respectively. The canonical quantization in the imaginary time
leads to the following commutation relations
\begin{equation}
[\Phi , \Phi^{*}]=1, \quad [\Psi , \Psi^{*}]=1,
\quad
[\Phi , \Psi^{*}]=0, \quad [\Psi , \Phi^{*}]=0 .
\label{e23}
\end{equation}
The Lagrangian can be presented in the form
\begin{equation}
{L}=\Phi^{*} \partial_y \Phi - \mu_P \Phi^{*} \Phi
+ \Psi^{*} \partial_y \Psi - \mu_O \Psi^{*} \Psi
+ V(\Phi,\Psi) .
\label{e21}
\end{equation}
Here $V(\Phi,\Psi)$ is the interaction term which does not
contain derivatives in $y$.
Then the Euclidean Hamiltonian is
\begin{equation}
H=-\left( \Phi^{*} \partial_y \Phi + \Psi^{*} \partial_y \Psi - L \right)
= -\mu_P \Phi^{*}\Phi - \mu_O \Psi^{*} \Psi + V(\Phi,\Psi) ,
\label{e24}
\end{equation}
where the interaction $V$ is
\begin{equation}
V(\Phi,\Psi)=
 i \lambda
\left( \Phi^{*}\Phi^{*}\Phi + \Phi^{*}\Phi\Phi
+2 \Psi^{*} \Psi \Phi +2 \Phi^{*} \Psi^{*} \Psi
- \Phi^{*} \Psi \Psi + \Psi^{*} \Psi^{*} \Phi \right).
\label{e24a}
\end{equation}

Comparing with the pomeron-odderon interaction introduced in
~\cite{BCV2} we observe that in our case all triple interactions are
imaginary but the relative sign of $P\to OO$ and $OO\to P$ transitions
is opposite. This guarantees that the $OO$ loops are have a different
sign as compared to $PP$ loops.  If one does the canonical phase
transformation \[ \Psi^*\to e^{-i\pi/4}\Psi^*,\ \ \Psi\to e ^{i\pi/4}\Psi , \]
then the two last terms in (\ref{e24a}) will appear with the same real coupling
constant $\lambda$ as in ~\cite{BCV2}. So our pomeron-odderon interaction
is fully equivalent to one in the latter, provided the three coupling constants
there have the same magnitude.

The relations \Ref{e23} allow one to interpret $\Phi^{*},\Psi^{*}$ as
the creation operators and $\Phi,\Psi$ as the annihilation operators.
The vacuum satisfies the condition $\Phi |0\> =\Psi|0\>=0$ and the
Fock space is considered as a full space of states created from
the vacuum by action of any number of $\Phi^{*}$ and $\Psi^{*}$.
The Schr\"odinger equation for the wave function of the system
$F(y,\Phi^{*},\Psi^{*})$ is
\begin{equation}
\frac{\pd F(y,\Phi^{*},\Psi^{*})}{\pd y}=-H F(y,\Phi^{*},\Psi^{*}) .
\label{schroed}
\end{equation}
Note that the original field $\Psi^{+}=-i\Psi^{*}$ has a meaning
of the creation operator of the odderon state.

\subsection{Equations of motion and comparison with the QCD equations
for amplitudes}

\noindent
The corresponding to \Ref{e24} differential equations of motion
for the field operators are
\begin{eqnarray}
{\pd{\Phi}}/{\pd y}&=[H,\Phi]= &\mu_P \Phi
-i{\lambda} \big( \Phi^2 - \Psi^2 \big)
-2i\lambda \big( \Phi^{*}\Phi + \Psi^{*}\Psi \big) , \nonumber \\
{\pd{\Psi}}/{\pd y}&=[H,\Psi]=&\mu_O \Psi -2i\lambda \Phi\Psi
-2i\lambda \big( \Psi^{*}\Phi + \Phi^{*}\Psi \big) , \nonumber \\
{\pd{\Phi}^{*}}/{\pd y}&=[H,\Phi^{*}]=& -\mu_P \Phi^{*}
+2i\lambda \big( \Phi^{*}\Phi + \Psi^{*}\Psi \big)
+i{\lambda} \big( \left(\Phi^{*}\right)^2 +\left(\Psi^{*}\right)^2 \big) ,
\nonumber \\
{\pd{\Psi}^{*}}/{\pd y}&=[H,\Psi^{*}]=&-\mu_O \Psi^{*}
+2i\lambda \big( \Psi^{*}\Phi - \Phi^{*}\Psi \big)
+2i\lambda \big( \Phi^{*}\Psi^{*} \big) .
\label{e44}
\end{eqnarray}
Note that the connection \Ref{e20} with the Hermitian conjugates
with respect to the standard scalar product is not conserved during
the evolution in $y$ since the Hamiltonian is not Hermitian and
the dynamics of the system is not unitary.

We are interested in these equations to compare them
in the fan approximation with the QCD. The fan equations are obtained
when all terms which contain two creation operators and one annihilation
operator are dropped from the Hamiltonian. In our case one has to consider
each of the equations \Ref{e44} without the last term in the parentheses.
Here we can introduce $\Phi=-i\chi$, $\Psi=-i\omega$
to get from two first equations
$$
\frac{\pd\chi}{\pd y}=\mu_P\chi+\lambda(-\chi^2+\omega^2),
$$
\begin{equation}
\frac{\pd\omega}{\pd y}=\mu_O\omega -2\lambda\chi\omega .
\label{faneq}
\end{equation}
One can see that these fan equations coincide in their form
with the fan equations in the QCD.

Indeed, the coupled equations for evolution of the odderon together
with the pomeron, derived in \cite{hiim} in the transverse coordinate
space are
$$
{\partial N(\vx,\vy;y) \over \partial y} =
{\abar \over 2\pi} \int d^2 z \;
{ (\vx - \vy)^2 \over (\vx - \vz)^2  (\vz - \vy)^2}
\left[
N(\vx,\vz;y) + N(\vz,\vy;y) - N(\vx,\vy;y) \right.
$$
\begin{equation}
\left.
- N(\vx,\vz;y)N(\vz,\vy;y)  +
O (\vx,\vz;y) O(\vz,\vy;y)
\right],
\label{e48}
\end{equation}
$$
{\partial O(\vx,\vy;y) \over \partial y} =
{\abar \over 2\pi} \int d^2 z \;
{ (\vx - \vy)^2 \over (\vx - \vz)^2  (\vz - \vy)^2}
\left[
O(\vx,\vz;y) + O(\vz,\vy;y) - O(\vx,\vy;y)
\right.
$$
\begin{equation}
\left.
- O(\vx,\vz;y)N(\vz,\vy;y)
- N (\vx,\vz;y) O(\vz,\vy;y)
\right]
\label{e49}
\end{equation}
where $N(\vx,\vy;y)$ and $O(\vx,\vy;y)$ are the $C$-even and $C$-odd
dipole amplitudes, respectively, symmetric and antisymmetric
in ${\bf x},{\bf y}$. Here $\abar=N_c\alpha_s/\pi$.

Following \cite{motyka} consider the translational invariant situation
when the amplitudes depend only on the difference ${\bf r=x-y}$.
Then passing to the momentum space
\begin{equation}
\chi(\vk,y) = \int {d^2 \vr \over 2 \pi r^2} N(\vr,y) \exp(-i \vk\vr),
\quad
\omega(\vk,y) = \int {d^2 \vr \over 2 \pi r^2} O(\vr,y) \exp(-i \vk\vr).
\end{equation}
we get the equations \cite{motyka}
$$
{\partial \chi(\vk,y) \over \partial y} =
-H^{BFKL}\chi -  \abar \chi^2 + \abar \omega^2,
$$
\begin{equation}
{\partial \omega(\vk,y) \over \partial y} =- H^{BFKL}\omega
- 2 \abar \chi\omega,
\label{e411}
\end{equation}
where $H^{BFKL}$ is the standard LL BFKL Hamiltonian.

For the local pomeron and odderon both $\chi$ and $\omega$ do not depend
on $\vk$ and depend only on rapidity $y$. Then the equations simplify to
$$
{\partial \chi(y) \over \partial y} =
\mu_P \chi(y) -  \abar \chi^2(y) + \abar \omega^2(y) ,
$$
\begin{equation}
{\partial \omega(y) \over \partial y} =
\mu_O \omega(y) -2 \abar \chi(y)\omega(y) ,
\label{e412}
\end{equation}
where the mass parameters $\mu_P$ and $\mu_O$ have a sense of
the pomeron and odderon intercepts minus $1$, respectively.
If one chooses $\lambda=\abar \equiv {N_c \alpha_s}/{\pi}$
then the equations \Ref{e412} completely coincide with our
equations \Ref{faneq}.

It is to be noted that the choice of real vertices for transitions
$O\to O+P$ and $O+P\to O$ in our original Lagrangian \Ref{e12a} was
dictated by the form of the equation for $\pd\omega/\pd y$. With the
imaginary vertices one would obtain factor $i$ in the right-hand part,
which could not be eliminated by rescaling of $\omega$, since the
equation is linear in it.

We acknowledge that the use of the one-dimensional description
without the transverse space dependence for the odderon field,
which is antisymmetric in $\vk$, is problematic. The comparison
of terms of the equations connected with local interactions can,
however, have a sense. It shows, at least, that coefficients before
different terms and their signs coincide in \Ref{e44} and \Ref{e412}.

Finally, the comparison of the fan equations fixes the coupling constant
for $\Phi^{*}\Psi\Psi$ interaction but not for the last term in \Ref{e24a}.
Here we choose the constant for the last term in \Ref{e12a} negative
imaginary (or, equivalently, the opposite signs before the two last terms
in \Ref{e24a}) to take into account the signature properties of odderon.
With the given choice of constants the simple loop constructed from two
odderon lines gives the positive correction to the pomeron propagator,
whereas the simple loop from one odderon line and one pomeron line
gives the negative correction both to the pomeron and odderon
propagators.

\subsection{Passage to the real Hamiltonian}

\noindent
The Lagrangian function \Ref{e12a} is to be accompanied by the form
of the coupling of the pomerons and odderons to the external particles.
We assume the eikonal form for the operators of creation of the initial
state and annihilation of the final state (analogous to ones taken
in \cite{braun1,braun2}):
\begin{equation}
F^{(in)}=1-e^{-ig_P^{(i)}\Phi^* + g_O^{{i}}\Psi^*} ,\quad
F^{(fin)}=1-e^{-ig_P^{(i)}\Phi - g_O^{{i}}\Psi}
\label{fin1}
\end{equation}
with possibly different coupling constants of interaction with
the projectile and the target. We choose $g_p$ with a negative sign
to later deal with well behaved expressions.

The Hamiltonian \Ref{e24} with interaction \Ref{e24a} is complex,
which makes practical use of it inconvenient. So one can pass
to a real Hamiltonian. One possibility (case A) to do it is
to introduce the Fock-Bargmann representation of operators
\begin{equation}
u \equiv i\Phi^{*} , \quad
v = \frac{\partial}{\partial u}\equiv -i\Phi , \ \ \quad
w \equiv i\Psi^{*} = -\Psi^{+} , \quad
z = \frac{\partial}{\partial w} \equiv -i\Psi .
\label{e53}
\end{equation}
In terms of these operators
\begin{equation}
H^{(A)}=-\mu_P u\frac{\pd}{\pd u}-\mu_O w\frac{\pd}{\pd w}
+\lambda u^2\frac{\pd}{\pd u}-\lambda u\frac{\pd^2}{\pd u^2}
-2\lambda w\frac{\pd^2}{\pd u\pd w}+2\lambda uw\frac{\pd}{\pd w}
+\lambda u\frac{\pd^2}{\pd w^2}+\lambda w^2\frac{\pd}{\pd u} .
\label{hampoa}
\end{equation}
The initial and final states become
\begin{equation}
F^{(in)}=1-e^{-g^{(i)}_P u - ig_O^{(i)} w} , \quad
F^{(fin)}=1-e^{g^{(f)}_P v - ig_O^{(f)} z}.
\label{e54a}
\end{equation}
So the coupling of the odderon to the projectile and the target
becomes pure imaginary.

Accordingly, our Hamiltonian is now real but the initial and final states
are not. To separate real and imaginary amplitudes one  has to separate
terms with even and odd numbers of odderons, that is having $C=+1$ and
$C=-1$. This can be done separating terms proportional to even or odd
powers $m=m_i+m_f$ of the product ${g^{(i)}_O}^{m_i}{g^{(f)}_O}^{m_f}$.

The alternative possibility (case B) is to only pass to imaginary
$\Phi$ and $\Phi^{*}$ and do not change $\Psi$ and $\Psi^*$:
\begin{equation}
u \equiv i\Phi^{*} , \quad
v = \frac{\partial}{\partial u}\equiv -i\Phi , \ \ \quad
w \equiv \Psi^{*} = i\Psi^{+} , \quad
z = \frac{\partial}{\partial w} \equiv \Psi .
\end{equation}
and to retain the form of the odderon coupling to the initial
and final states. The alternative Hamiltonian is
\begin{equation}
H^{(B)}=-\mu_Pu\frac{\pd}{\pd u}-\mu_O w\frac{\pd}{\pd w}
+\lambda u^2\frac{\pd}{\pd u}-\lambda u\frac{\pd^2}{\pd u^2}
-2\lambda w\frac{\pd^2}{\pd u\pd w}+2\lambda uw\frac{\pd}{\pd w}
-\lambda u\frac{\pd^2}{\pd w^2}-\lambda w^2\frac{\pd}{\pd u} .
\label{hampob}
\end{equation}
As compared to the previous case the signs of the two last terms
will be opposite. However, it is trivial to find out that this
possibility is equivalent to the canonical transformation
$\Psi\to i\Psi,$ $\Psi^*\to -i\Psi^*$
and in the end gives the same result for the amplitudes.

%%3
\section{Evolution in rapidity}
\subsection{Basic equations and methods for solution}

\noindent
If the initial state in the Schr\"odinger picture at rapidity zero
is given by
\begin{equation}
|{y=0}\>=F_{y=0}(\Phi^{*},\Psi^{*})|0\>,
\end{equation}
then at rapidity $y$ it will be given by
\begin{equation}
|y\>=F_y(\Phi^{*},\Psi^{*})|0\>
\end{equation}
with the evolution to rapidity $y$ given by the equation
\begin{equation}
\frac{\partial F_y}{\partial y}=-HF_y .
\label{evol}
\end{equation}
The amplitude for the transition from the state $F_{y=0}|0\>$
to a state $\<0|F^{(fin)}$ at rapidity $y$ will be given by the
matrix element
\begin{equation}
i{\cal A}(y)=
\<0| F^{(fin)}(\Phi,\Psi)
e^{-Hy}
F_{y=0}(\Phi^{*},\Psi^{*}) |0\> .
\label{e51}
\end{equation}

We take the Hamiltonian in the real form \Ref{hampoa} with
the initial and final states in the form \Ref{e54a}.
The amplitude becomes
$$
i{\cal A}(y)=
\<0| \Big(1-e^{g_P^{(f)}v -ig_O^{(f)}z}\Big)
e^{-Hy} \Big(1-e^{-g_P^{(i)}u - ig_O^{(i)}w}\Big) |0\>
$$
\begin{equation}
=-\<0|
\, e^{g_P^{(f)}\frac{\pd}{\pd u} - ig_O^{(f)} \frac{\pd}{\pd w}}
\, F_y(u,w) |0\> ,
\label{e55}
\end{equation}
where
\begin{equation}
F_y(u,w)=e^{-Hy}(1-e^{-g_P^{(i)}u -ig_O^{(i)}w})
\label{e56}
\end{equation}
is the result of the evolution of the initial state to rapidity $y$.

Applying the derivatives we find
\begin{equation}
i{\cal A}(y)=-F_y(u,w)\Big|_{u=g^{(f)}_P, w=-ig^{(f)}_O},
\label{e57}
\end{equation}
that is the amplitude is obtained by just substituting $u$ and $w$
in the evolved $F(u,w)$ with the coupling constant for the target.
To find the propagator one evidently has to take the terms linear
in these coupling constants (or in $u$ or $w$). So the pomeron and
odderon propagators are obtained as
\begin{equation}
P(y)=\frac{\pd F(u,w)}{\pd u}\Big|_{u=0,w=0},
\ \ O(y)=\frac{\pd F(u,w)}{\pd w}\Big|_{u=0,w=0}
\label{props}
\end{equation}
with the odderon propagator actually carrying extra factor $i$.

Passing to the possible methods to solve the evolution equation
\Ref{evol} with the Hamiltonian given by \Ref{hampoa} or
\Ref{hampob} we can find two basic alternatives. The first one is
to study the Hamiltonian and find its eigenvalues. This is a difficult
problem already without odderon, as demonstrated by earlier studies.
With only the pomeron the Hamiltonian is non-Hermitian and acting on
a complex field variable. Fortunately, in this case it could be transformed
to a Hermitian Hamiltonian, which allowed to approximately find its
ground state at very small coupling \cite{jengo}. Later a formalism
has been developed to find all eigenvalues from the initial Hamiltonian
in the complex variable \cite{braun3}. Inclusion of the odderon
substantially aggravates the situation. Now the initial non-Hermitian
Hamiltonian depends on two complex variables. Its analysis requires much
more efforts and transformation to a Hermitian Hamiltonian becomes
hardly possible.

So we are left with the second alternative, which is
to directly integrate the evolution equation numerically. This procedure
turned out quite efficient for the pure pomeron model at physically
reachable energies \cite{braun1}, although of course it cannot be
stretched to infinite energies when the groundstate eigenvalue becomes
essential. So we turn to direct integration of \Ref{evol} starting
from some initial condition at $y=0$.

One may consider two possibilities for the choice of variables for
the wave function. One may develop $F_y(u,w)$ in power expansion
in $u$ and $w$
\begin{equation}
F_y(u,w)=\sum_{n=0,m=0}g_{nm}(y)u^nw^m.
\label{powers}
\end{equation}
This form has a nice physical interpretation. Term with $u^nw^m$
corresponds to a state with $n$ pomerons and $m$ odderons. So expansion
\Ref{powers} immediately gives the pomeron-odderon content of the wave
function. However, convergence of this expansion is not guaranteed and
the cut series obviously has a very bad behaviour at large $u$ and $w$.
So, as we shall see, the applicability of this method is severely
restricted to small values of $y$ and the coupling constant $\lambda$.

Another possibility actually employed in \cite{braun1} is to take the
initial function $F_{y=0}(u,w)$ on the lattice $(u_i,w_j)$ and evolve
$F$ on this lattice. In this case the evolution equation itself is
responsible for the high $y$ and $\lambda$ behaviour. As a result this
method can be applied  in a wide region of $y$ and $\lambda$.

In the next subsections we consider these numerical methods
in more detail.

\subsection{Evolution by power expansion}

\noindent
Using \Ref{powers}  one finds from \Ref{hampoa}
$$
H F_y(u,w)=\sum_{n=0,m=0}g_{nm}(y)
\Big(-\mu_Pnu^nw^m-\mu_Omu^nw^m
+\lambda nu^{n+1}w^m-\lambda n(n-1)u^{n-1}w^m
$$
\begin{equation}
-2\lambda nmu^{n-1}w^m
+2\lambda mu^{n+1}w^m+\lambda m(m-1)u^{n+1}w^{m-2}
+\lambda nu^{n-1}w^{m+2}\Big) .
\end{equation}
Combining these terms as a coefficient before $u^nw^m$ one finds
\begin{equation}
H F_y(u,w)=\sum u^nw^m f_{nm}(y),\ \ \ f_{nm}(y)=Hg_{nm}(y) ,
\end{equation}
where at $n\ge 2$ and $m\ge 2$
$$
f_{nm}=-\mu ng_{nm}-\mu_Omg_{nm}
+\lambda(n-1)g_{n-1,m}-\lambda n(n+1)g_{n+1,m}
$$
\begin{equation}
-2\lambda(n+1)mg_{n+1,m}
+2\lambda mg_{n-1,m}+\lambda (m+1)(m+2)g_{n-1,m+2}
+\lambda (n+1)g_{n+1,m-2}
\label{fnm}
\end{equation}
and for smaller values of $n$ and $m$
$$f_{00}=0,\ \ f_{10}=-\mu_Pg_{10}-2\lambda g_{20}+2\lambda g_{02},
\ \ f_{01}=-\mu_Og_{01}-2\lambda g_{11} ,
$$
$$f_{11}=-(\mu_P+\mu_O)g_{11}-6\lambda g_{21}+2\lambda g_{01}+6\lambda g_{03} ,
$$
$$f_{20}=-2\mu_P g_{20}+\lambda g_{10}-6\lambda g_{30}+2\lambda g_{12}
$$
$$f_{21}=-(2\mu_p+\mu_O)g_{21}+3\lambda g_{11}-12\lambda g_{31}+6\lambda g_{13} ,
$$
$$f_{12}=-(\mu_P+2\mu_O)g_{12}-10\lambda g_{22}
-4\lambda g_{02}+12\lambda g_{04}+2\lambda g_{20} ,
$$
\begin{equation}
f_{02}=-2\mu_O g_{02}-4\lambda g_{12} +\lambda g_{10} .
\end{equation}
Note that if the initial $g_{nm}$ are given at $n,m\le N$ then one has
to require that in \Ref{fnm} terms with $g_{nm}$ outside this domain
should be put to zero.

The full Hamiltonian describes creation and absorption of pomerons
and odderons so that the theory includes loops constructed of both.
To study the influence of loops one can consider a reduced fan
Hamiltonian which describes fan diagrams and does not contain loops.
Denoting the successive eight terms in \Ref{hampoa} or \Ref{hampob}
as $(1),(2),\dots ,(8)$ one finds that the fan Hamiltonian
contains terms $(1)+(2)+(3)+(6)+(8)$ and does not contain the rest.

So in case A the fan Hamiltonian is
\begin{equation}
H^{(A)}_{fan}=-\mu_Pu\frac{\pd}{\pd u}-\mu_Ow\frac{\pd}{\pd w}
+\lambda (u^2+w^2)\frac{\pd}{\pd u} +2\lambda uw\frac{\pd}{\pd w}
\label{hfan}
\end{equation}
and in $H^{(B)}_{fan}$ the term with $w^2$ will have the opposite
sign.

For the fan diagrams in the power representation
instead of \Ref{fnm} we find for $n,m\ge 2$
\begin{equation}
f_{nm}^{fan}=-(n\mu_P+m\mu_O)g_{nm}
+\lambda (n-1+2m)g_{n-1,m}+\lambda (n+1)g_{n+1,m-2}
\label{fanfnm}
\end{equation}
and for $n<2$ or $m<2$ with $\mu_O=0$
$$f_{00}=0,\ \ f_{10}=-\mu_Pg_{10},
\ \ f_{01}=0,f_{11}=-\mu_P g_{11}+2\lambda g_{01} ,
$$
$$f_{20}=-2\mu_P g_{20}+\lambda g_{10},
\ \ f_{21}=-2\mu_P g_{21}+3\lambda g_{11} ,
$$
\begin{equation}
f_{12}=-\mu_P g_{12}+4\lambda g_{02}+2\lambda g_{20},
\ \ f_{02}=+\lambda g_{10} .
\end{equation}

\subsection{Evolution by points. Composite fields}

\noindent
The method similar to one used in the previous papers \cite{braun1,braun2}
is to use the evolution equation as it is, choosing some initial function
at a set of points $u_0,\dots, u_N, w_0,\dots,w_N$ and evolving it with
rapidity $y$ by the Runge-Kutta method. As a result of this numerical
exercise one gets $F_y(u_0,...u_N|w_0,...w_N)$. This method worked for
the pomeron evolution without the odderon in a very wide choice
of the model parameters.

Note, however, that this method requires a very particular choice of variables
$u$ and $w$. In the Gribov model studied in \cite{braun1,braun2} the wave function
is an analytic functions in $u$. Obviously, any analytical function
can be completely determined by its values on some ray in the complex
plane. However, one experience drawn from those papers was that the initial
condition for the evolution has to be imposed on the positive real axis
$i\Phi^*=u>0$, otherwise the numerical evolution process breaks down.
The reason may be in that in practice one has to limit the values of $\Phi^*$
(or $u$). Convergence requires the wave function to fall at large values of the argument
and this is possible only for  a very particular direction in the complex $\Phi^*$-plane.
As a result one has obligatory to pass from $\Phi^*$ to $u=i\Phi^*$ and start the evolution
from positive $u$.

With the odderon included we have the same situation with two variables
$\Phi^*$ and $\Psi^*$ and  from the start it is not clear which values
for these variables in their complex plane allow convergence. We have
immediately discovered that with the choice $u=i\Phi^*$ and $w=i\Psi^*$
evolution from the positive axes for both variables is impossible,
leading to divergence at quite small values of $y<1$.

So we are bound to seek for different variables in which evolution becomes realizable.
As a hint one may consider a model (which will be the subject of our subsequent
paper) in which the fake odderon does not possess the negative signature
and moreover has the same intercept as the pomeron, their difference reduced only
to the $C$-parity. In such a model the choice of variables for evolution is trivial,
since it follows this choice for the model with only the pomeron.

Based on this hint we introduce  composite fields as linear
combinations of the pomeron and odderon fields. We define
$$
\phi = \frac{\Phi + i \Psi}{\sqrt{2}} ,
\quad
\phi^* = \frac{\Phi^* - i \Psi^*}{\sqrt{2}} ,
$$
\begin{equation}
\tilde \phi = \frac{\Phi - i \Psi}{\sqrt{2}} ,
\quad
\tilde \phi^* = \frac{\Phi^* + i \Psi^*}{\sqrt{2}} .
\label{e31}
\end{equation}
Their commutation relations are
$$
\left[  \phi , \phi^* \right] = 1 , \quad
\left[ \tilde \phi , \tilde \phi^* \right] = 1 ,
$$
\begin{equation}
\left[ \phi , \tilde \phi \right] =
\left[ \phi , \tilde \phi^* \right] =
\left[ \phi^* , \tilde \phi \right] =
\left[ \phi^* , \tilde \phi^* \right] = 0 ,
\label{e32}
\end{equation}
which shows that $\phi$ and $\tilde\phi$ are independent
dynamical variables.

If one takes the interaction Hamiltonian as a sum of two
interactions of the Gribov model \Ref{e11} depending
on different composite variables and with an appropriate
coupling constants one finds
\begin{equation}
V_0 = i {\sqrt{2}}\,{\lambda}
\left(\phi^{*}\phi\phi + \phi^{*}\phi^{*}\phi \right)
+
i {\sqrt{2}}\,{\lambda}
\left(\tilde\phi^{*}\tilde\phi\tilde\phi
+\tilde\phi^{*}\tilde\phi^{*}\tilde\phi \right)
\label{e33}
\end{equation}
or in terms of elementary fields
\begin{equation}
V_0(\Phi,\Psi)=
+ i \lambda \left( \Phi^{*}\Phi\Phi + \Phi^{*}\Phi^{*}\Phi
+2 \Psi^{*}\Psi\Phi + 2 \Phi^{*}\Psi^{*}\Psi
- \Phi^{*}\Psi\Psi - \Psi^{*}\Psi^{*}\Phi \right).
\label{e34}
\end{equation}
Comparing with the true interaction $V$ with the odderon signature
taken into account we find
\beq
V=V_0+V_1,\eeq
where
\beq
V_1=2i\lambda\Psi^*\Psi^*\Phi=
-i\frac{\lambda}{\sqrt{2}}\Big(\phi^*
-\tilde{\phi}^* \Big)^2 (\phi+\tilde{\phi}).
\eeq
So the full Hamiltonian in terms of composite fields is
\beq
H=H_0+V_0+V_1,
\eeq
where the free part (in the physical case $\mu_O=0$) is
\beq
H_0=-\frac{1}{2}\mu_P (\phi^* +\tilde{\phi}^*)(\phi+\tilde{\phi}).
\eeq

Following the experience with only the pomeron field we pass
to real variables introducing
\begin{equation}
\phi^*=-iu , \quad
\phi=i\frac{\pd}{\pd u} =iv , \quad
\tilde{\phi}^*=-iw , \quad
\tilde{\phi}=\frac{\pd}{\pd w} =iz .
\end{equation}
We stress that these $u$ and $w$ are not the old $u$ and $w$
in \Ref{e53}. We denote them so only to economize in notations.
Then the final Hamiltonian in variables $u$ and $w$ is found as
$$
H=-\frac{1}{2}\mu_P (u+w)\Big(\frac{\pd}{\pd u}+\frac{\pd}{\pd w}\Big)
$$
$$
-\lambda\sqrt{2}\Big(u\frac{\pd^2}{\pd u^2}-u^2\frac{\pd}{\pd u}
+w\frac{\pd^2}{\pd w^2}-w^2\frac{\pd}{\pd w}\Big)
$$
\begin{equation}
-\frac{\lambda}{\sqrt{2}}(u-w)^2\Big(\frac{\pd}{\pd u}+\frac{\pd}{\pd w}\Big).
\label{newh}
\end{equation}

This Hamiltonian is symmetric in variables $u$ and $w$. So once
the initial wave function $F(u,w)$ is (anti)symmetric in $u$ and $w$
it will preserve this property in the evolution in rapidity.

It is remarkable that if one takes the initial function $F(u,v)$ on the
positive axis of both $u$ and $v$ then its evolution in rapidity become
possible in a rather wide region of parameters $\mu_P$ and $\lambda$
and also to rather high values of the rapidity. The results of this
evolution for different sets of $(\mu_P,\lambda)$ are reported in the
next Section.

Note that the pomeron and odderon fields are expressed as
\begin{equation}
\Phi^+=\Phi^*=\frac{\phi^* +\tilde \phi^*}{\sqrt{2}},
\quad
\Psi^+=-i\Psi^*=\frac{\phi^*-\tilde \phi^*}{\sqrt{2}}
\label{e60}
\end{equation}
or in terms of $u$ and $w$
\begin{equation}
\Phi^+=-i\frac{u+w}{\sqrt{2}},\ \ \Psi^+=-i\frac{u-w}{\sqrt{2}}
\label{e61}
\end{equation}
and the evolution equation for the wave function takes the form
\Ref{evol} with the Hamiltonian \Ref{newh}. To start evolution
in terms of $\phi$, $\tphi$ we choose \Ref{e61}, dropping the common
factor $(-i)$, as the initial state $F_{y=0}(u,w)$ for the pomeron
and odderon, respectively. The expressions for the propagators become
\begin{equation}
P(y)=\frac{1}{\sqrt{2}}
\left(\frac{\partial}{\partial u} + \frac{\partial}{\partial w} \right)
F(u,w)\Big|_{u=w=0},
\ \ O(y)=-\frac{1}{\sqrt{2}}
\left(\frac{\partial}{\partial u} - \frac{\partial}{\partial w} \right)
F(u,w)\Big|_{u=w=0}
\label{e62}
\end{equation}
instead of \Ref{props}. Since the exchange of $u$ and $w$ is
the $C$-transformation, the $C$-symmetry of the Hamiltonian assures
that the wave function remains symmetrical under $u \leftrightarrow w$
for the pomeron and antisymmetrical for the odderon at any $y$.

Furthermore, $C$-symmetry guarantees that the symmetrical
and antisymmetrical parts of the wave function evolve independently
that allows one to choose the sum $u\sqrt{2}$ of \Ref{e61} as the initial
state and to find both propagators from \Ref{e62} in one numerical
simulation.

%%4
\section{Numerical evolution: results}
\subsection{Power expansion}

\noindent
Evolution by power expansion works only for small values of $\mu_P$ and
$\lambda$. In our calculation we took $\mu=0.1$ and $\lambda =0.01$
and $0.03$. The evolution equations in all cases were solved by the second
order Runge-Kutta method with the precision 10000 points for $\Delta y=1$.
The evolution breaks down at larger $\lambda$. For the chosen $\lambda$
the evolution works approximately until $y=30$ for $\lambda=0.01$ and until
$y=20$ for  $\lambda=0.03$. The actual interval of $\lambda$ when
the evolution works is clear from the figures presenting the results.
Note that for technical reasons our figures are labeled by
values of $e=-\mu$ and $a=-\lambda$.

\newpage
%\vspace*{0.2 cm}
{\bf The propagators}

\begin{figure}[h!p]
\begin{center}
\includegraphics[width=7.2 cm]{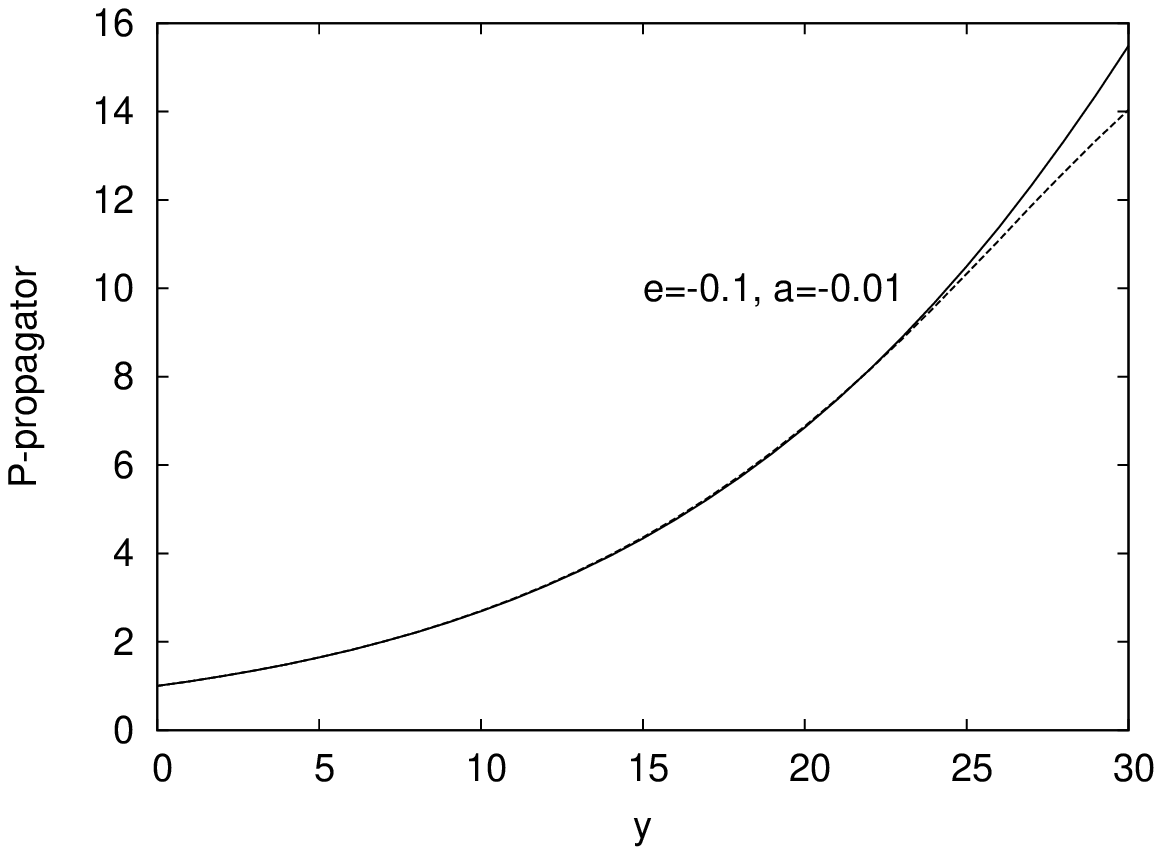}
\includegraphics[width=7.2 cm]{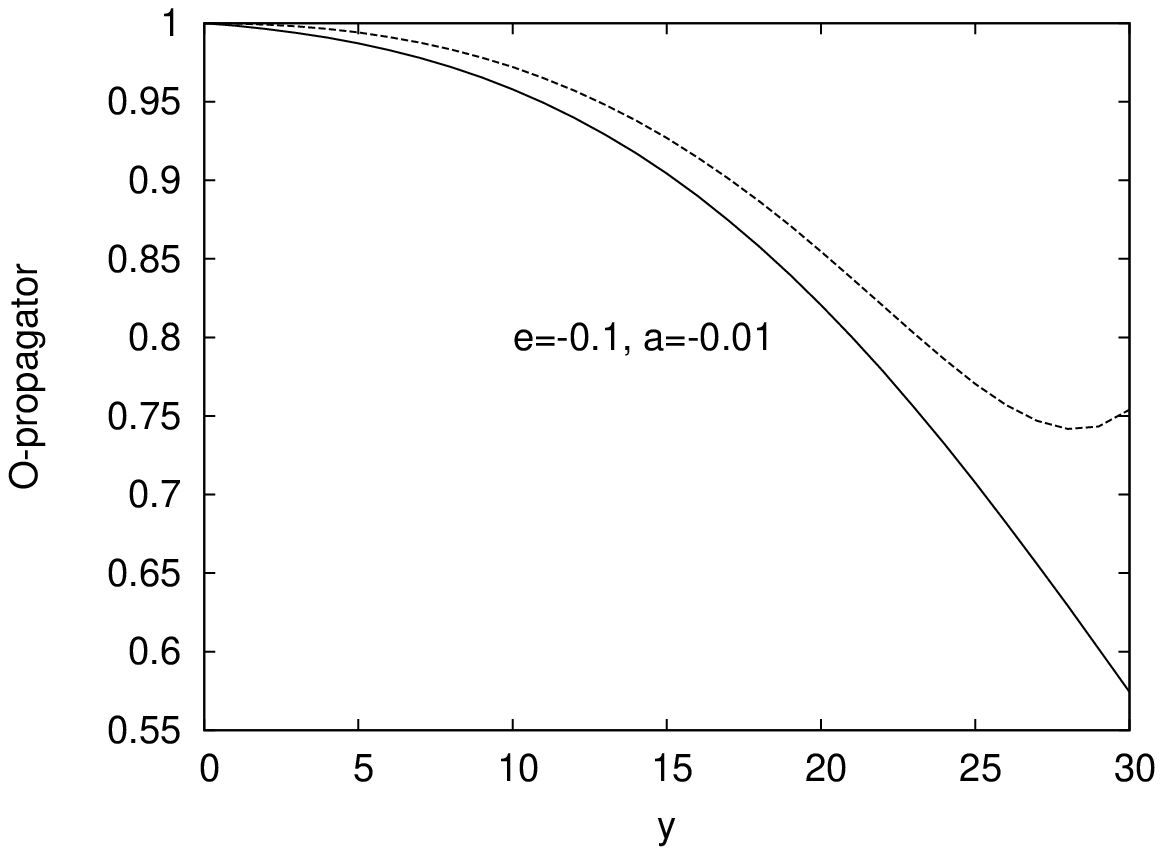}
\includegraphics[width=7.2 cm]{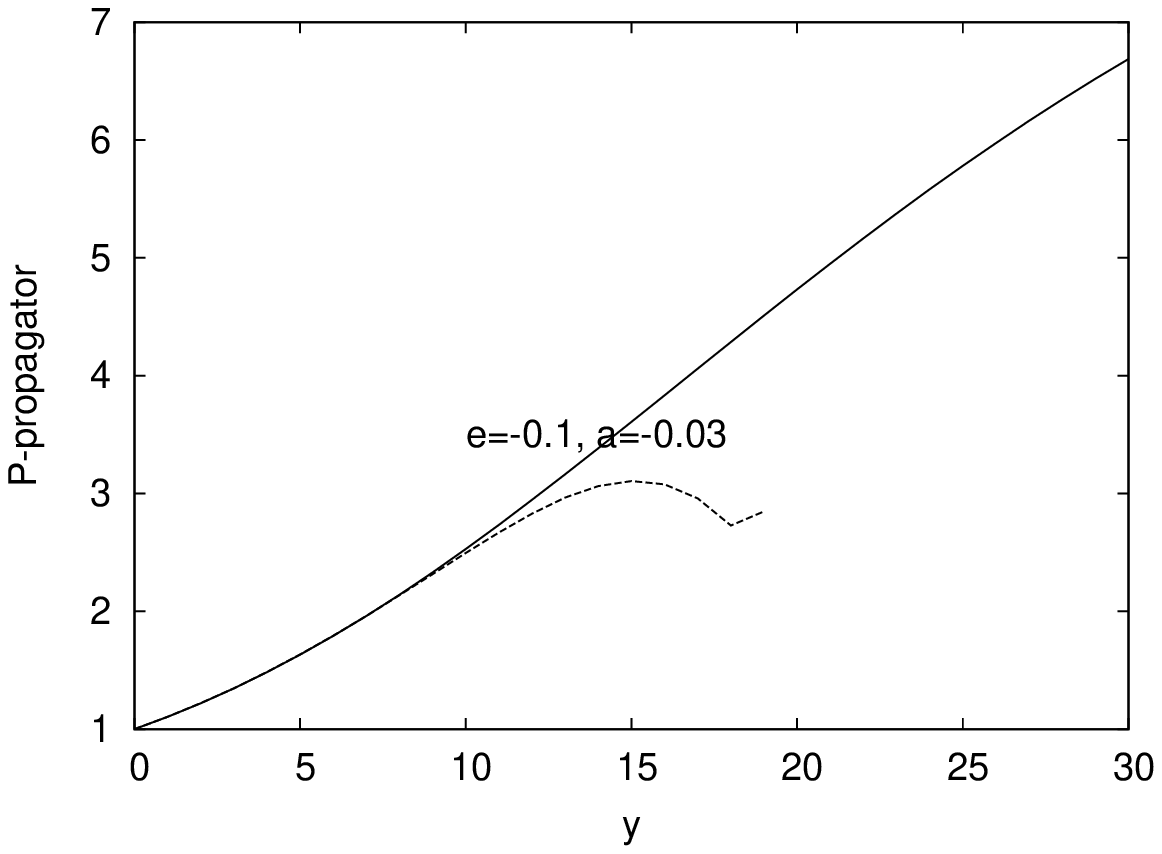}
\includegraphics[width=7.2 cm]{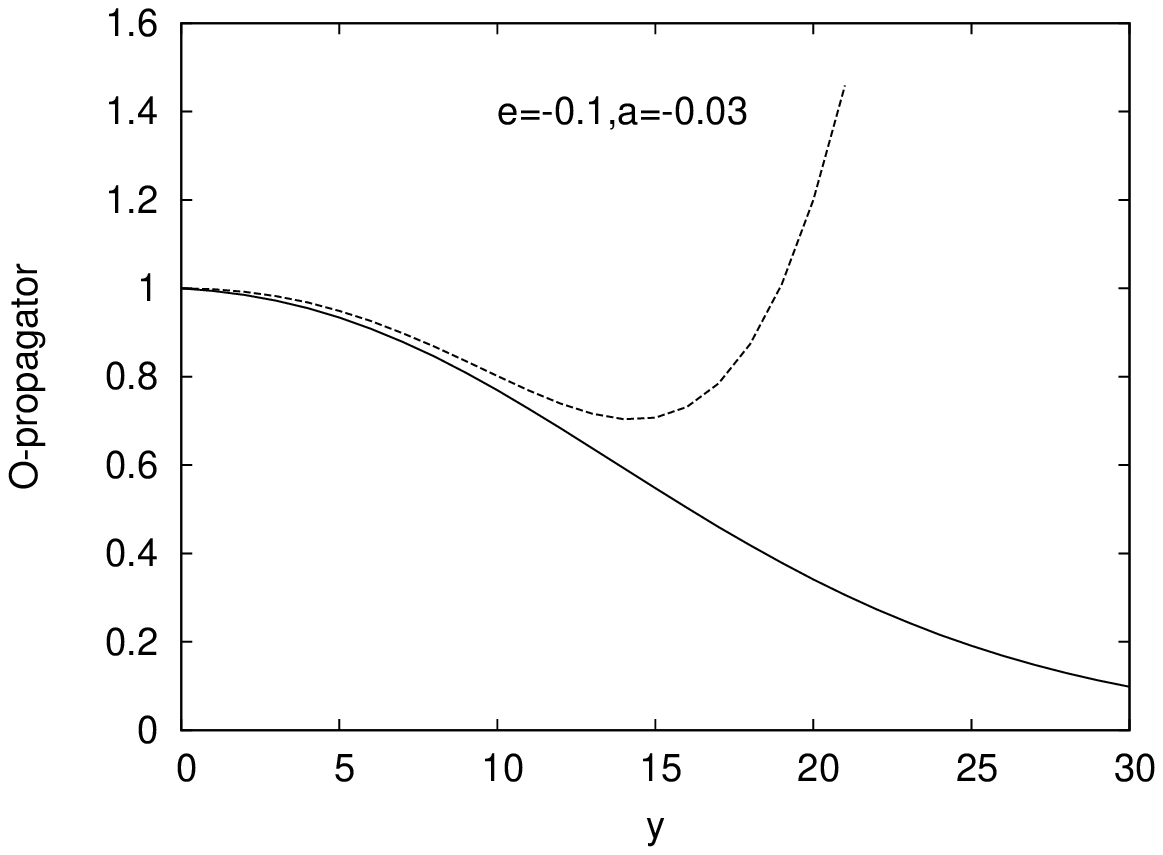}
\vskip -0.2cm
\caption{Propagators of the pomeron (left panels) and odderon (right panels)
as calculated by power expansion (dashed curves) and point evolution (solid curves).
In all panels $\mu=0.1$. In upper panels $\lambda=0.01$, in the lower $\lambda=0.03$.}
\label{fig6}
\end{center}
\vskip -0.5cm
\end{figure}

\noindent
The pomeron propagator $P(y)$ calculated by power expansion for
$\mu=0.1$ is shown in Fig.~\ref{fig6} by upper curves in the left upper
panel for $\lambda=0.01$ and the left lower panel for $\lambda=0.03$.
To compare and see the precision the lower curves present our far more
precise results obtained by point evolution. In the right panels
we similarly present the odderon propagator for the same values of $\mu$
and $\lambda$. For the odderon the lower curves were obtained by power
expansion and the upper ones by point evolution. One observes that
the power expansion gives satisfactory results at smaller values
of $\lambda$ and rapidity. With the growth of both its precision worsens
notably and at $\lambda=0.03$ its convergence breaks down already
at $y\sim 15$.

The influence of loops and the odderon in the whole region $0<y<30$
will be illustrated later, in Fig. \ref{fig1}
based on the precise calculations by the point-like evolution.

\vspace*{0.2 cm}
{\bf The pA amplitude}

\noindent
The $pA$ amplitude may be $C$-even and $C$-odd and depend on the couplings
of the pomeron and odderon to the proton in the projectile and target.
We take $g^{(i)}_P=g^{(f)}_P=1$ and for the odderon we consider two
cases $|g_O^{(i)}|=1$ and $g_O=0$. In the second case all the influence
of the odderon is reduced to loops.
The coupling constant $\lambda$ was again taken as $0.01$
and $0.03$.

\begin{figure}[ht]
\begin{center}
\includegraphics[width=7.2 cm]{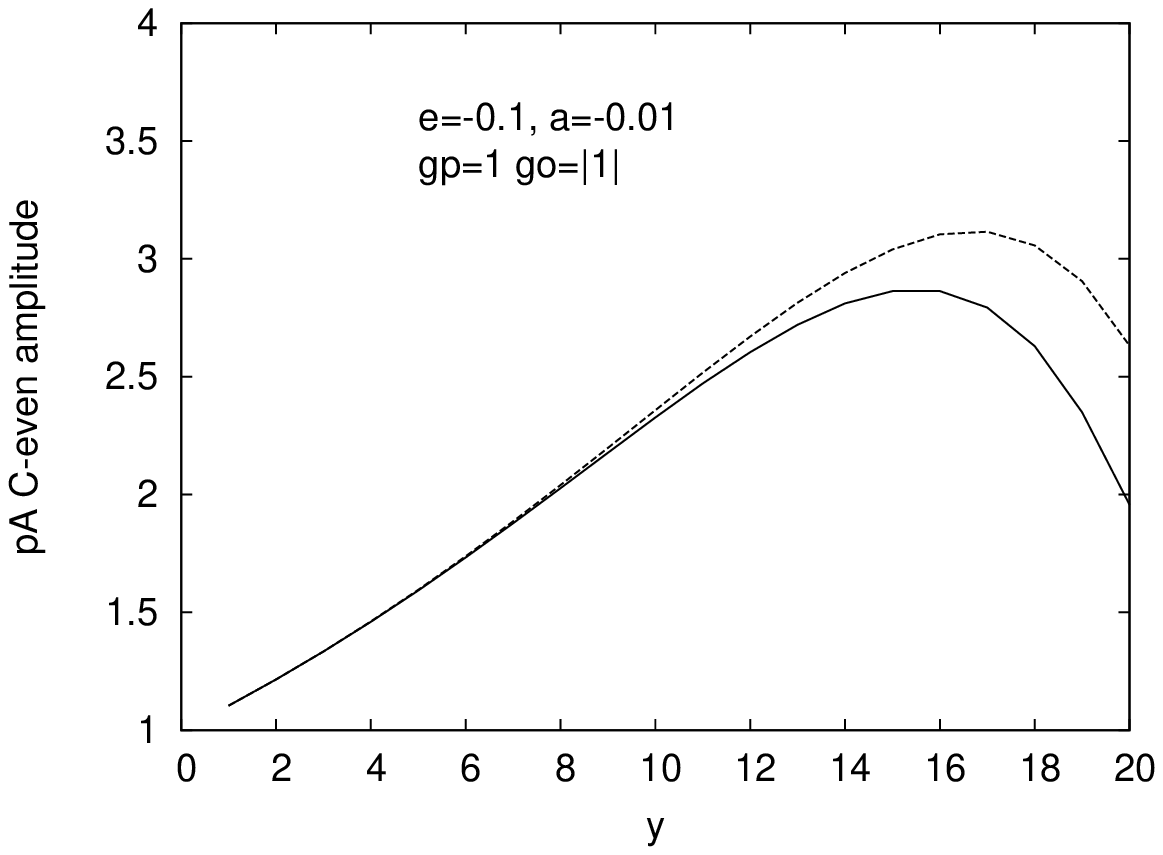}
\includegraphics[width=7.2 cm]{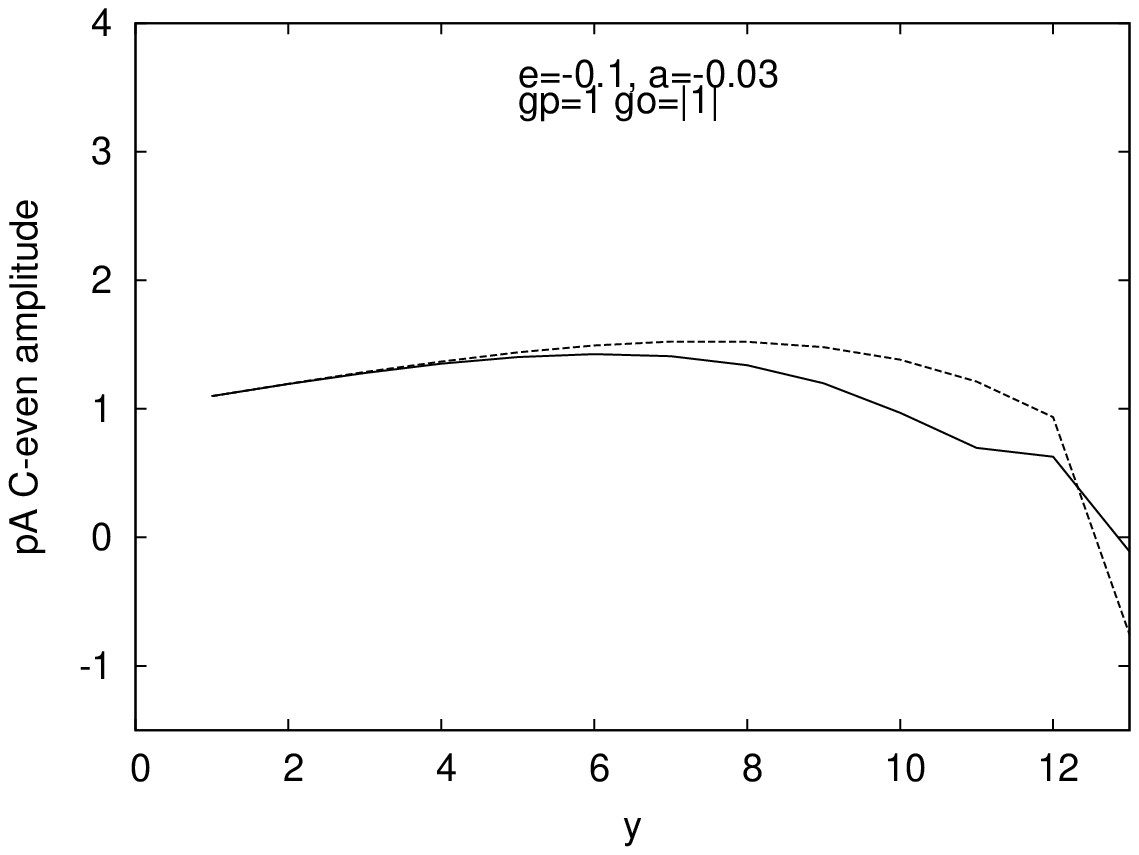}
\caption{pA C-even amplitudes for at different rapidities
for $\mu_p=0.1 $, $\lambda=0.01, 0.03$ and $g_P=|g_O|=1$.
The upper curves correspond to summation of the fan diagrams.}
\label{fig3}
\end{center}
\end{figure}

\begin{figure}[ht]
\begin{center}
\includegraphics[width=7.2 cm]{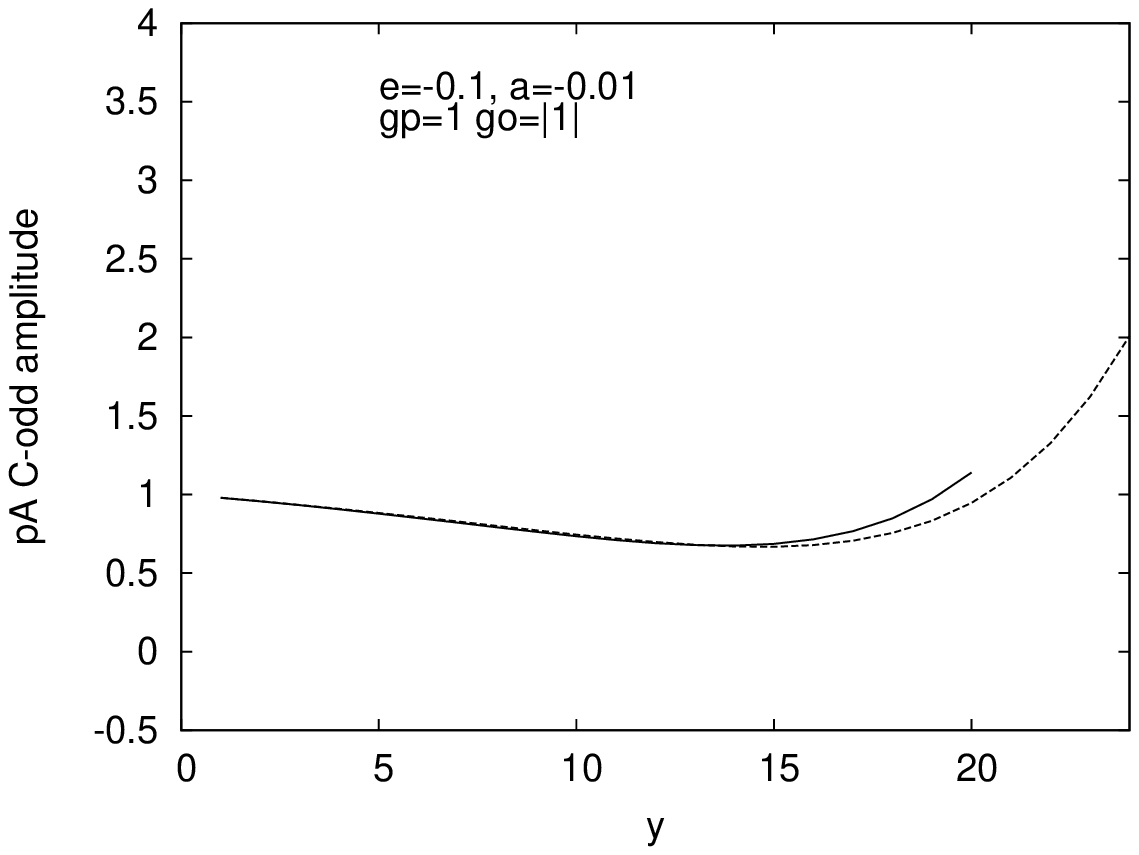}
\includegraphics[width=7.2 cm]{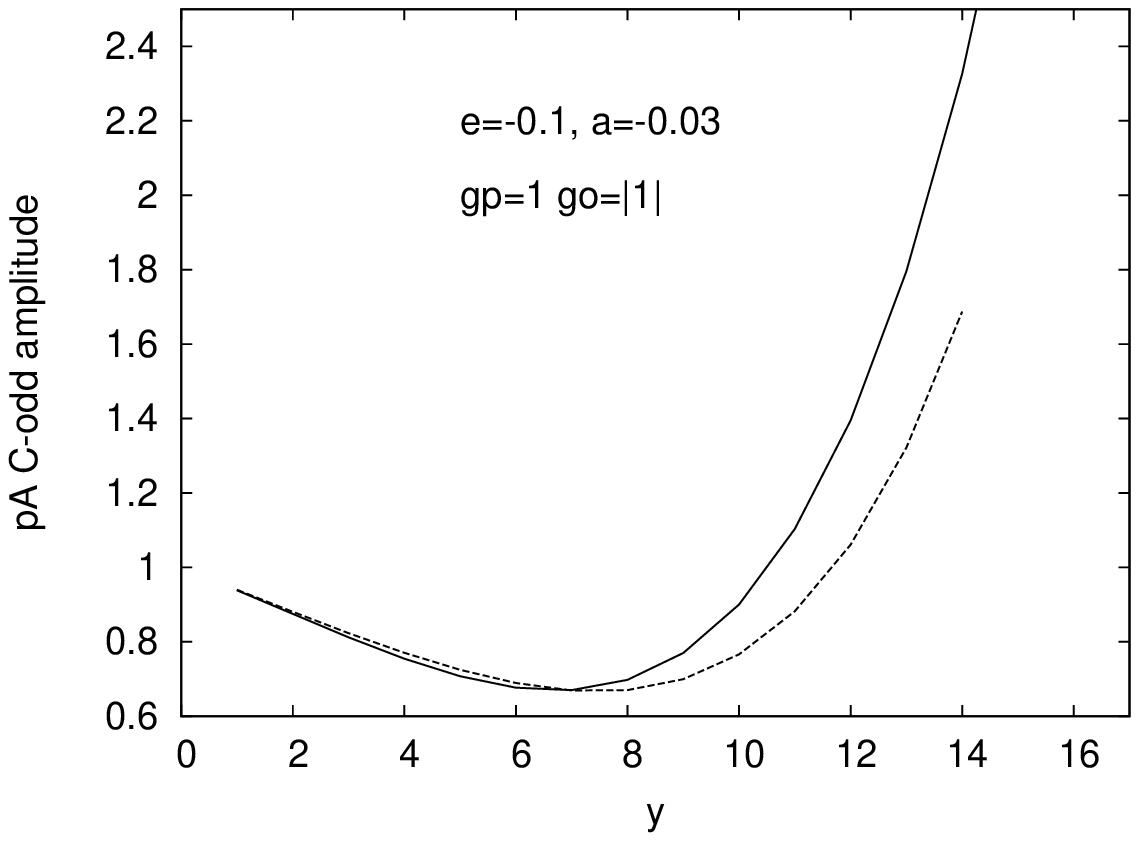}
\caption{pA C-odd amplitudes at different rapidities
for $\mu_p=0.1$, $\lambda=0.01, 0.03$ and $g_P=|g_O|=1$.
The lower curves correspond to summation of the fan diagrams.}
\label{fig4}
\end{center}
\end{figure}

\begin{figure}[h!]
\begin{center}
\includegraphics[width=7.2 cm]{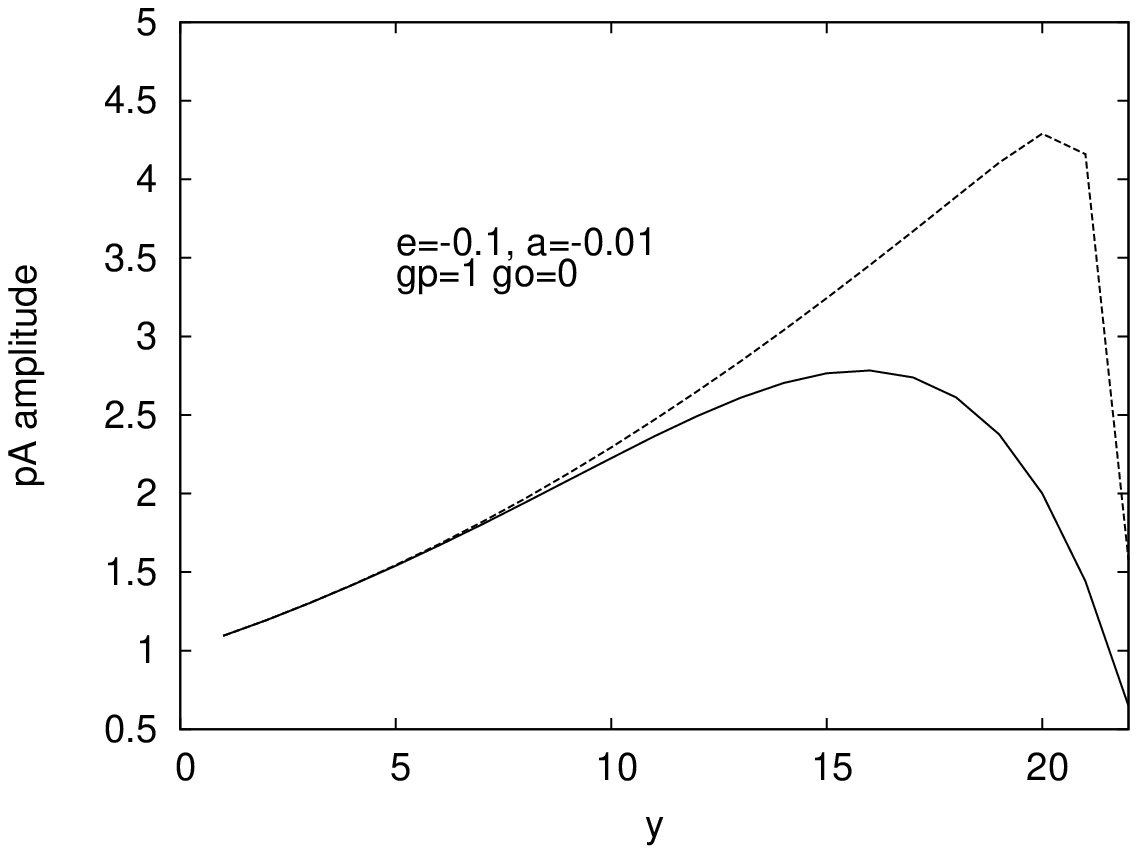}
\includegraphics[width=7.2 cm]{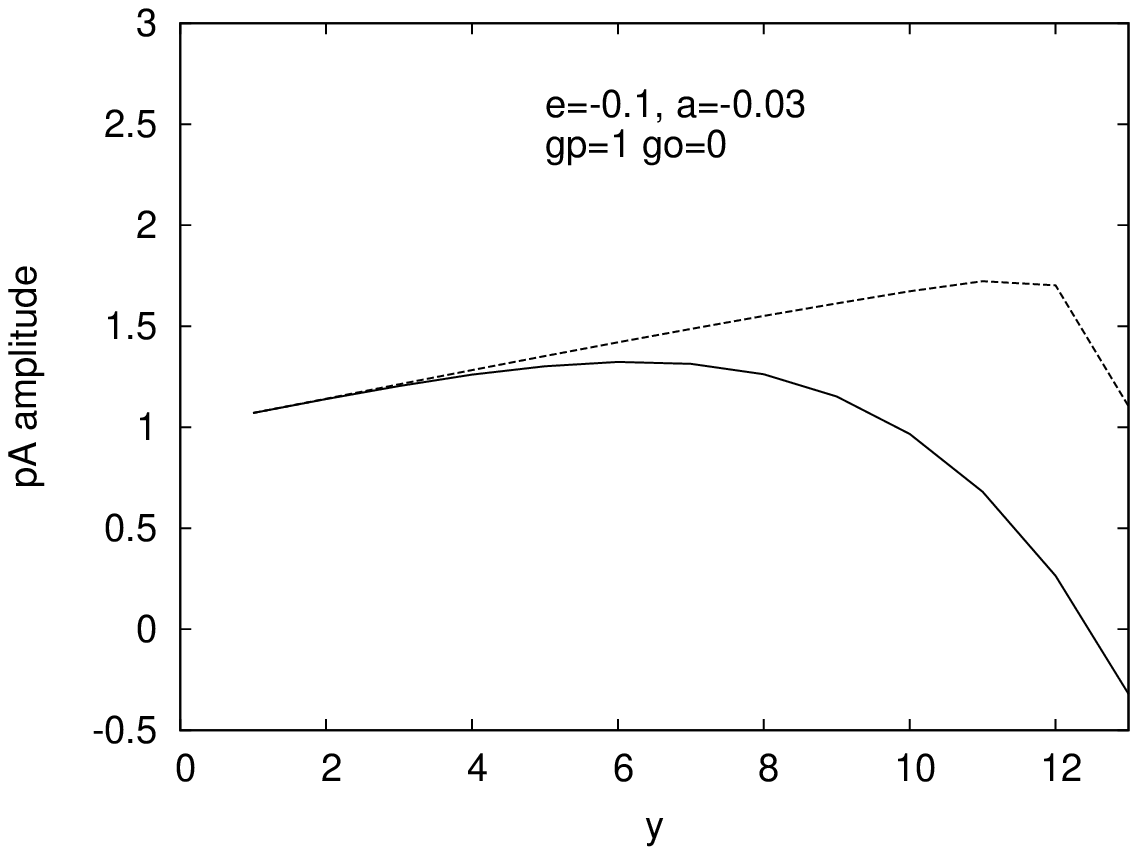}
\caption{pA C-even amplitudes at different rapidities
for $\mu_p=0.1$, $\lambda=0.01, 0.03$ and $g_P=1, g_O=0$ (lower curves).
The upper curves corresponds to summation of the fan diagrams.}
\label{fig5}
\end{center}
\end{figure}

The $C$-even and $C$-odd amplitudes are presented in Figs. \ref{fig3}
and Fig.~\ref{fig4}. In Fig.~\ref{fig5} we show
the even amplitude with the odderon coupling equal to zero, when the whole
influence of the odderon reduces to its loop contribution. In all these
pictures the $pA$ amplitude is shown together with the predictions from
fan diagrams. As one observes in the $C$-even amplitude the combined effect
of the pomeron and odderon loops  is the same as in the propagators: they
diminish the amplitudes, just as without odderon. The situation with
the $C$-odd amplitude is not clear, since it is quite small and the loop
effects become visible at rapidities probably outside the region
of convergence, which is observed at $y>5$ for $\lambda=0.03$. 

\subsection{Evolution by points}

\noindent
As mentioned, unlike evolution by powers, evolution of Eq.~\Ref{evol}
on the $(u,w)$ lattice admits
wide areas in $y$, $\mu$ and $\lambda$ using different steps in $y$.
The latter depend on the values of $\lambda$.
For $\lambda\leq 0.05$ satisfactory results were obtained with $N=400$
points in the interval $(0,20)$ for both $u$ and $w$ and 2000 points
for $\Delta y=1$. However, for $\lambda=0.1$ we had to raise the precision
in $y$ to take 20000 points for $\Delta y=1$. With $\lambda=1$ one has
to take 100000 points for $\Delta y=1$ with the corresponding rise
in the processor time. Raising the number of points in $u$ and $w$
to $N=500$ changes both propagators only by few percents. Below we report
on the calculated pomeron and odderon propagators, for which we took
symmetric and antisymmetric initial conditions $(u\pm w)/\sqrt{2}$,
respectively. We do not report our results for $pA$ amplitudes, since
their behaviour more or less follows the one found above via the evolution
by powers.

\vspace*{0.2 cm}
{\bf Pomeron and odderon propagators}

We start with our results for the pomeron at small values
of $\mu$ and $\lambda$, where also power expansion gives converging results
previously shown in Fig. \ref{fig6}.

\begin{figure}[h!p]
\begin{center}
\includegraphics[width=7.2 cm]{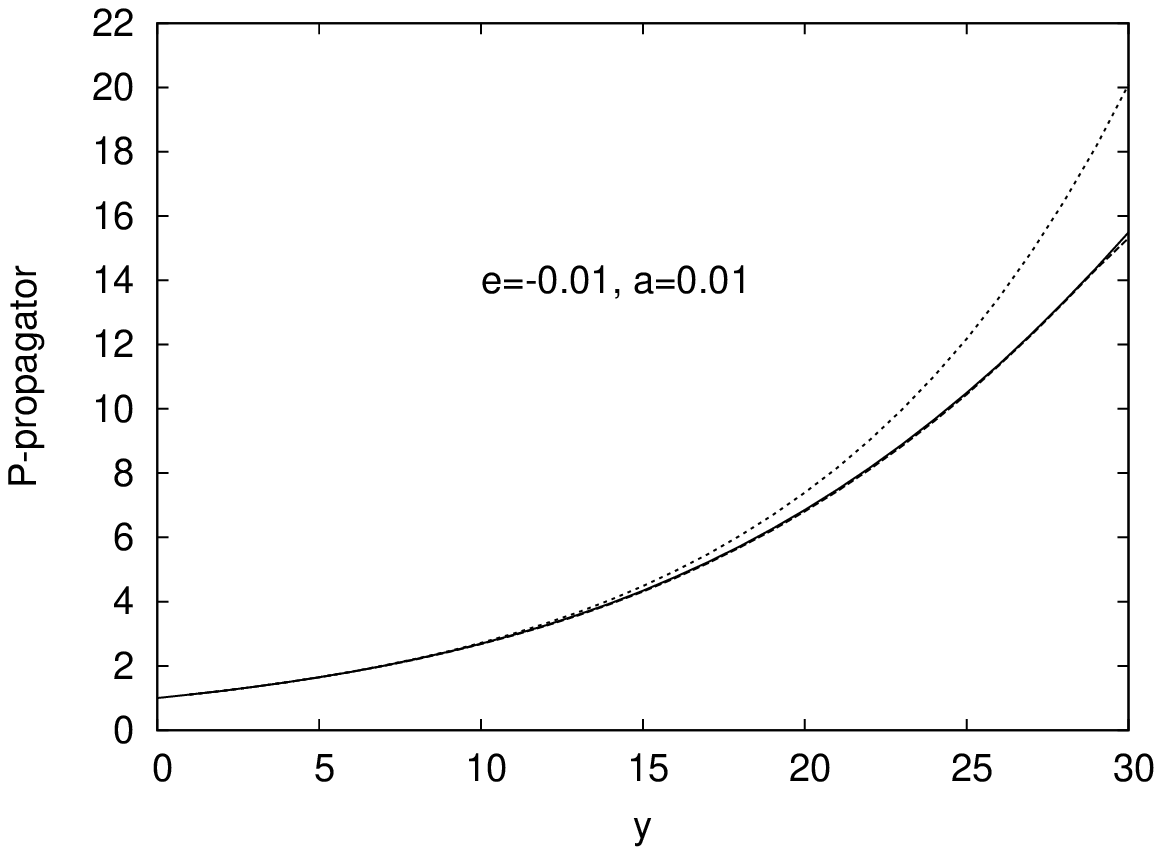}
\includegraphics[width=7.2 cm]{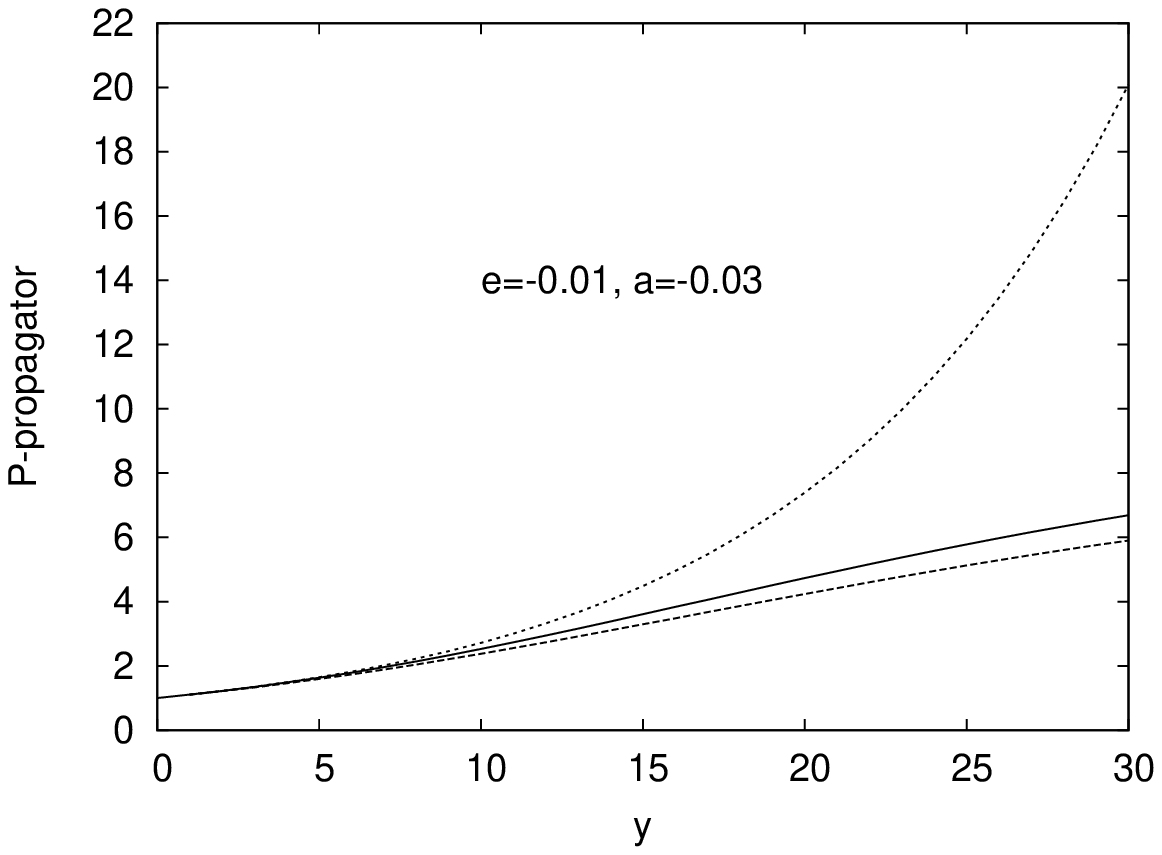}
\vskip -0.2cm
\caption{The pomeron propagators
as functions of rapidity for $\mu=0.1$, $\lambda=0.01$ (left panel)
and $0.03$ (right panel). Our results with both the pomeron and odderon loops
are shown in the middle curve, those with only the pomeron loops
in the bottom curve and without loops in the upper curve.}
\label{fig1}
\end{center}
\vskip -0.5cm
\end{figure}

In Fig. \ref{fig1} the solid (middle) curves show the pomeron propagator
with $\mu=0.1$ and $\lambda=0.01$ (left panel) and $0.03$ (right panel).
The bottom curves show the results obtained without the odderon
(only pomeron loops) and the upper curves show the propagator without
any loops. One concludes that the odderon loops somewhat enhance
the propagator, which is natural as they bear the opposite sign.

\begin{figure}[h!p]
\begin{center}
\includegraphics[width=7.2 cm]{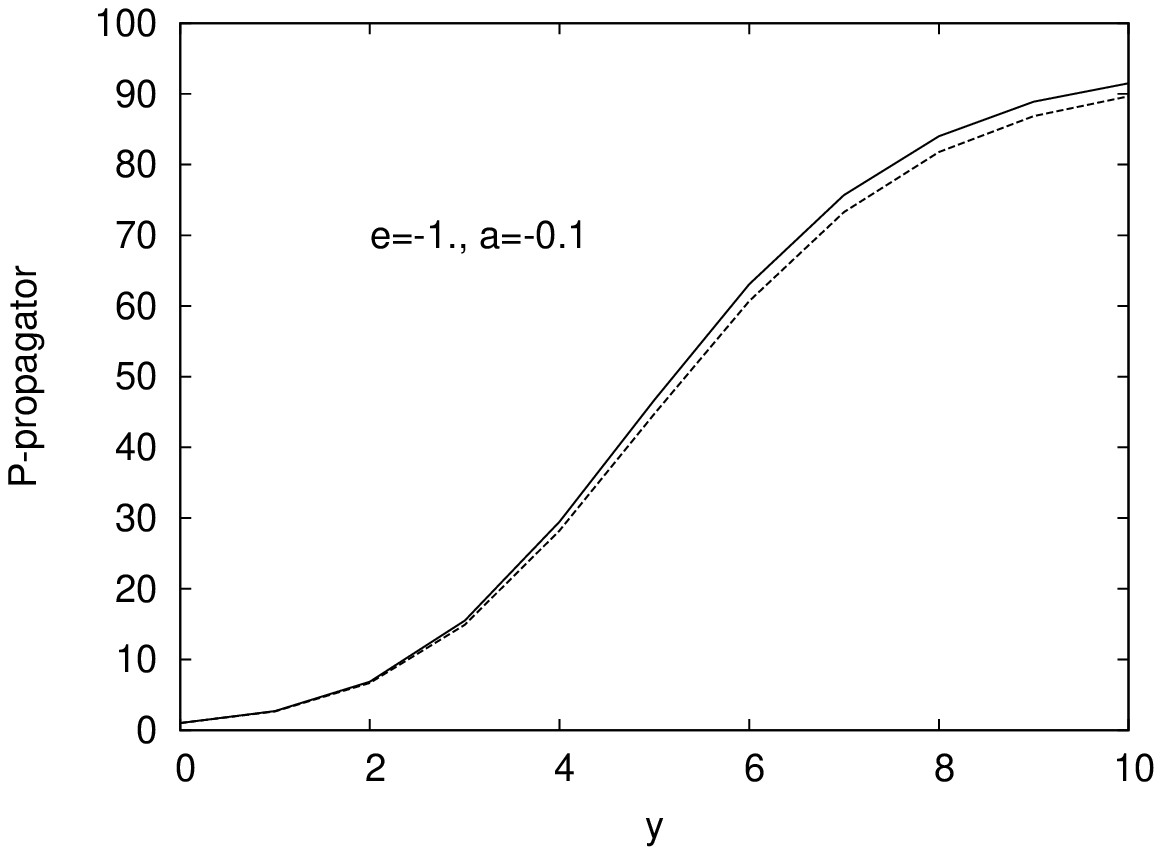}
\includegraphics[width=7.2 cm]{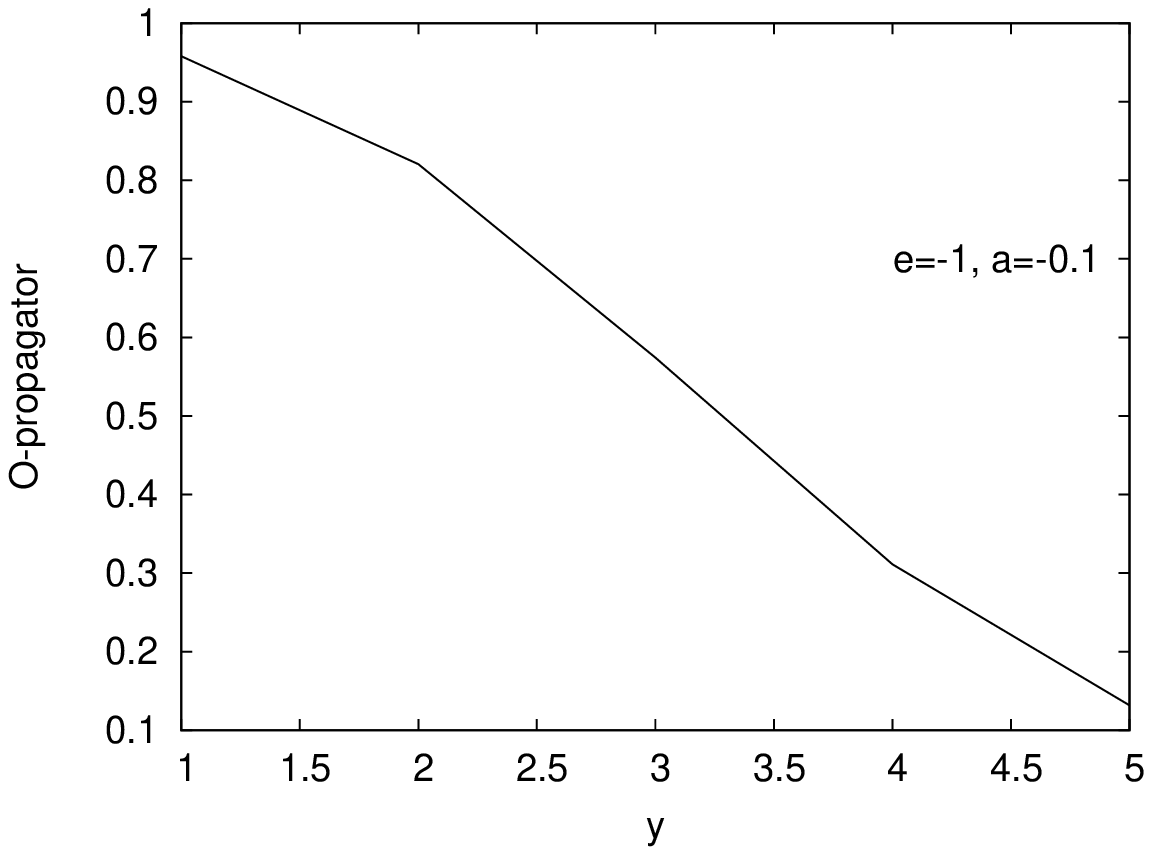}
\caption{The solid curves show pomeron (left panel) and odderon (right panel) propagators
as functions of rapidity for $\mu=1$, $\lambda=0.1$. The dashed curve in the left panel
shows the pomeron propagator in absence of the odderon (only pomeron loops).}
\label{fig7}
\end{center}
\end{figure}

\begin{figure}[h!p]
\begin{center}
\includegraphics[width=7.2 cm]{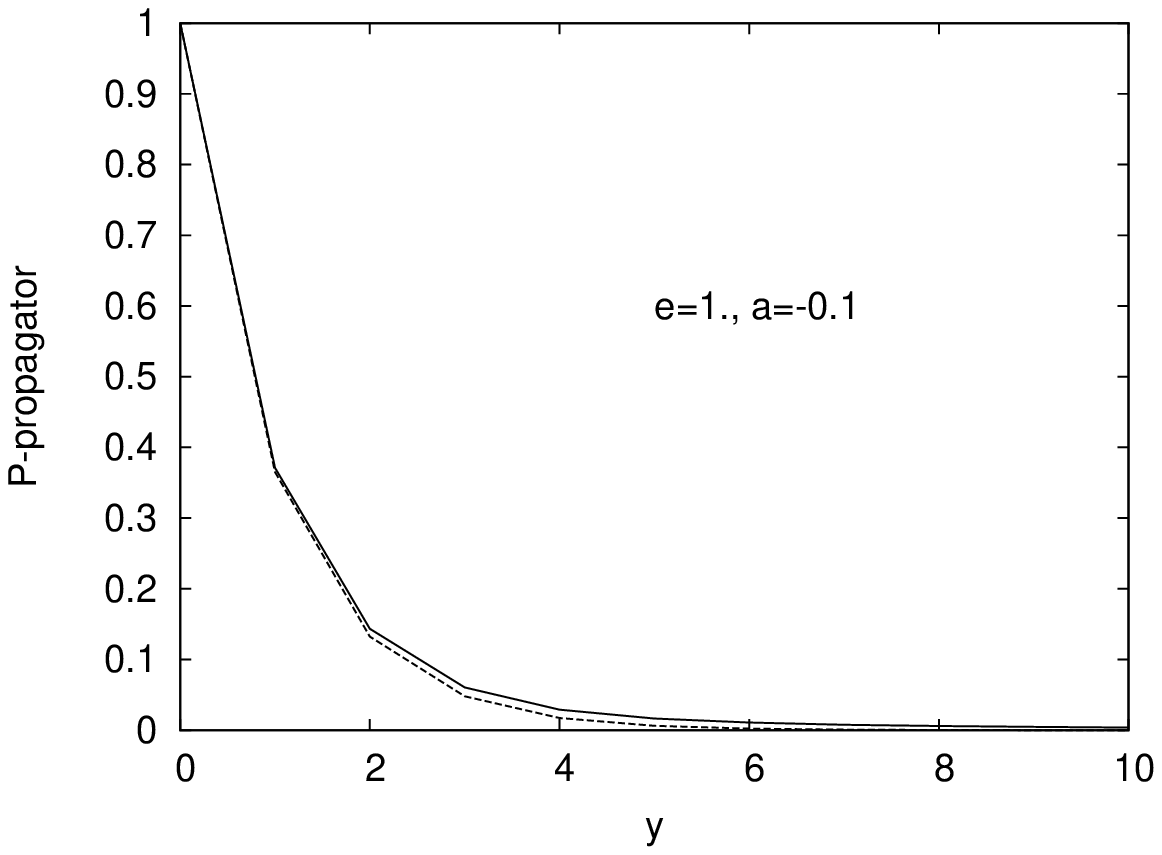}
\includegraphics[width=7.2 cm]{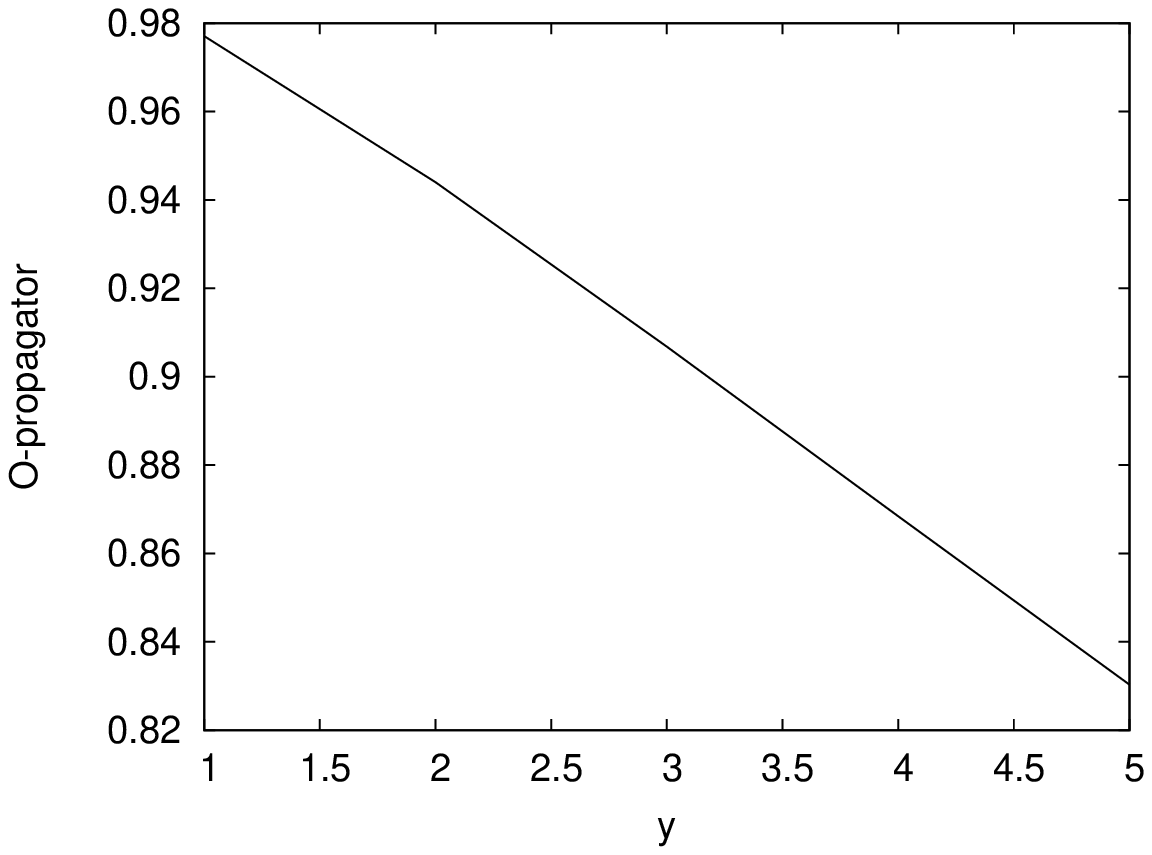}
\caption{The solid curves show pomeron (left panel) and odderon (right panel) propagators
as functions of rapidity for $\mu=-1$, $\lambda=0.1$. The dashed curve in the left panel
shows the pomeron propagator in absence of the odderon (only pomeron loops).}
\label{fig8}
\end{center}
\end{figure}

\begin{figure}[h!p]
\begin{center}
\includegraphics[width=7.2 cm]{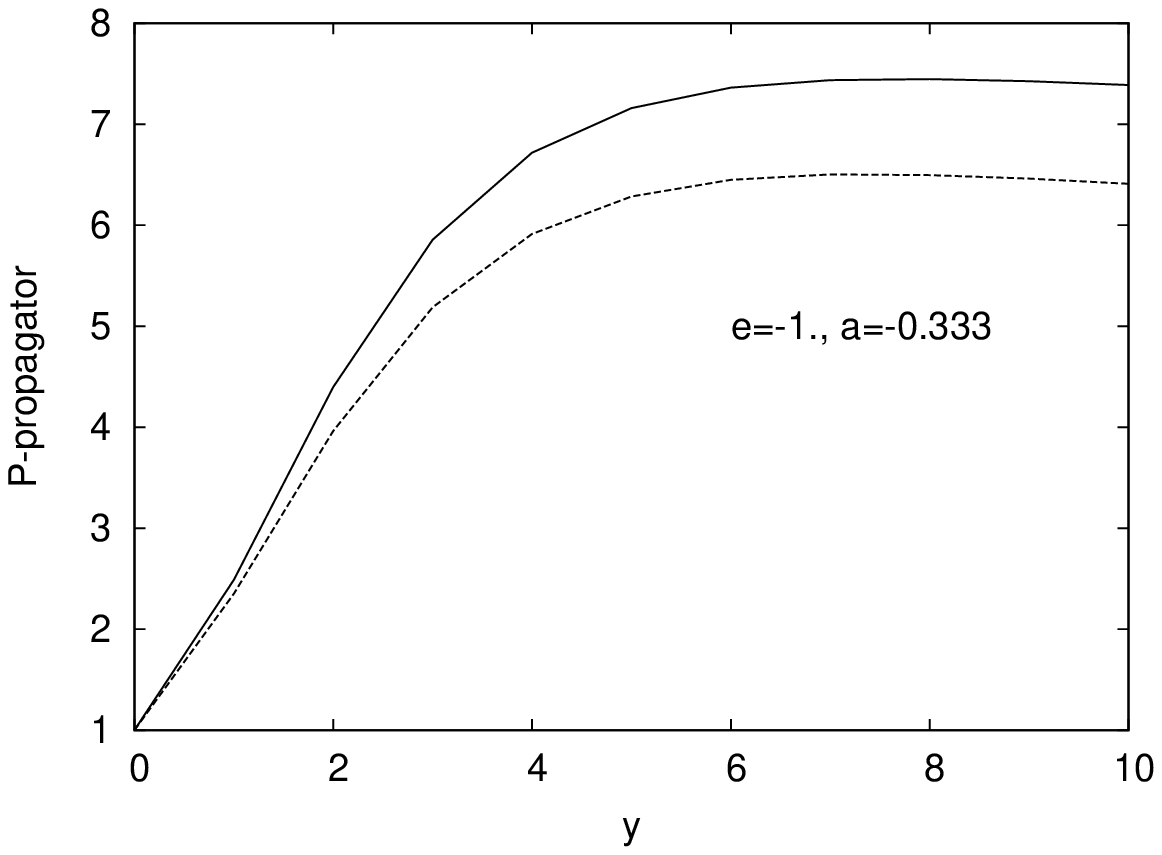}
\includegraphics[width=7.2 cm]{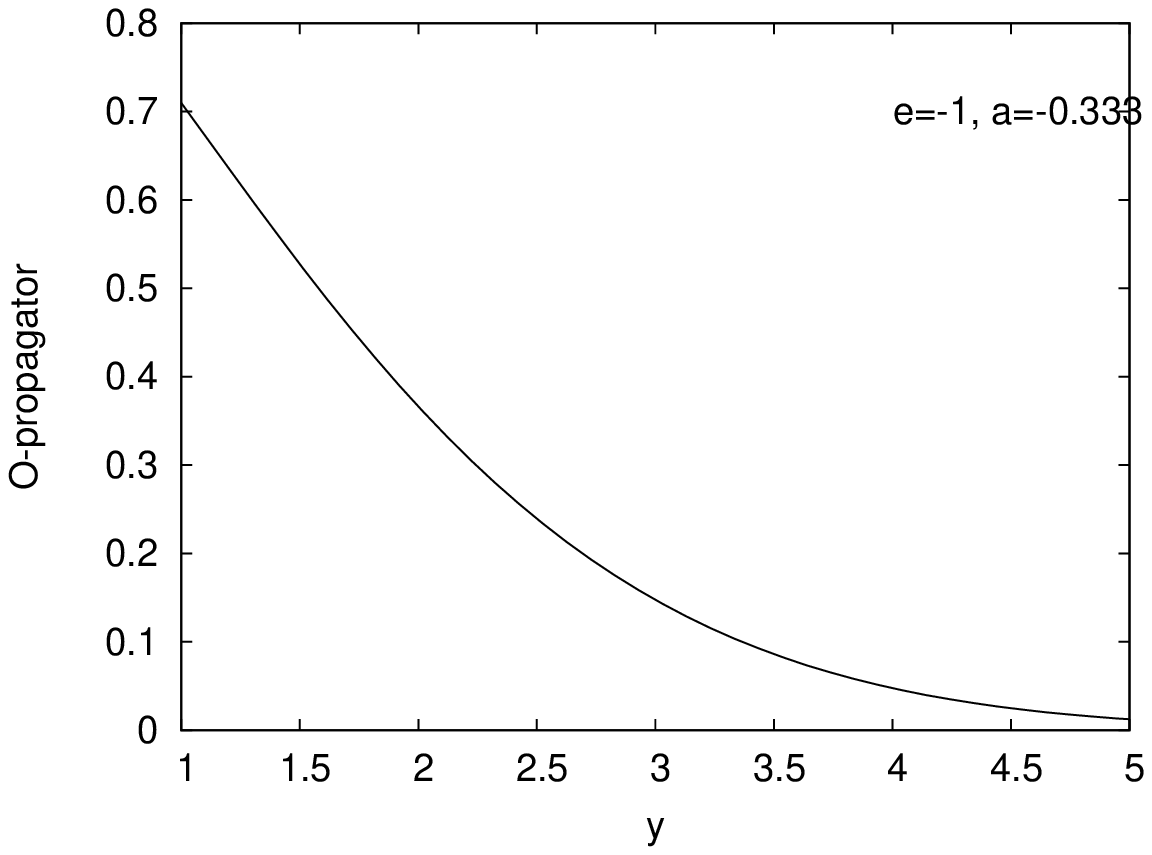}
\caption{The solid curves show pomeron (left panel) and odderon (right panel) propagators
as functions of rapidity for $\mu=1$, $\lambda=1/3$. The dashed curve in the left panel
shows the pomeron propagator in absence of the odderon (only pomeron loops).}
\label{fig9}
\end{center}
\end{figure}

\begin{figure}[h!p]
\begin{center}
\includegraphics[width=7.2 cm]{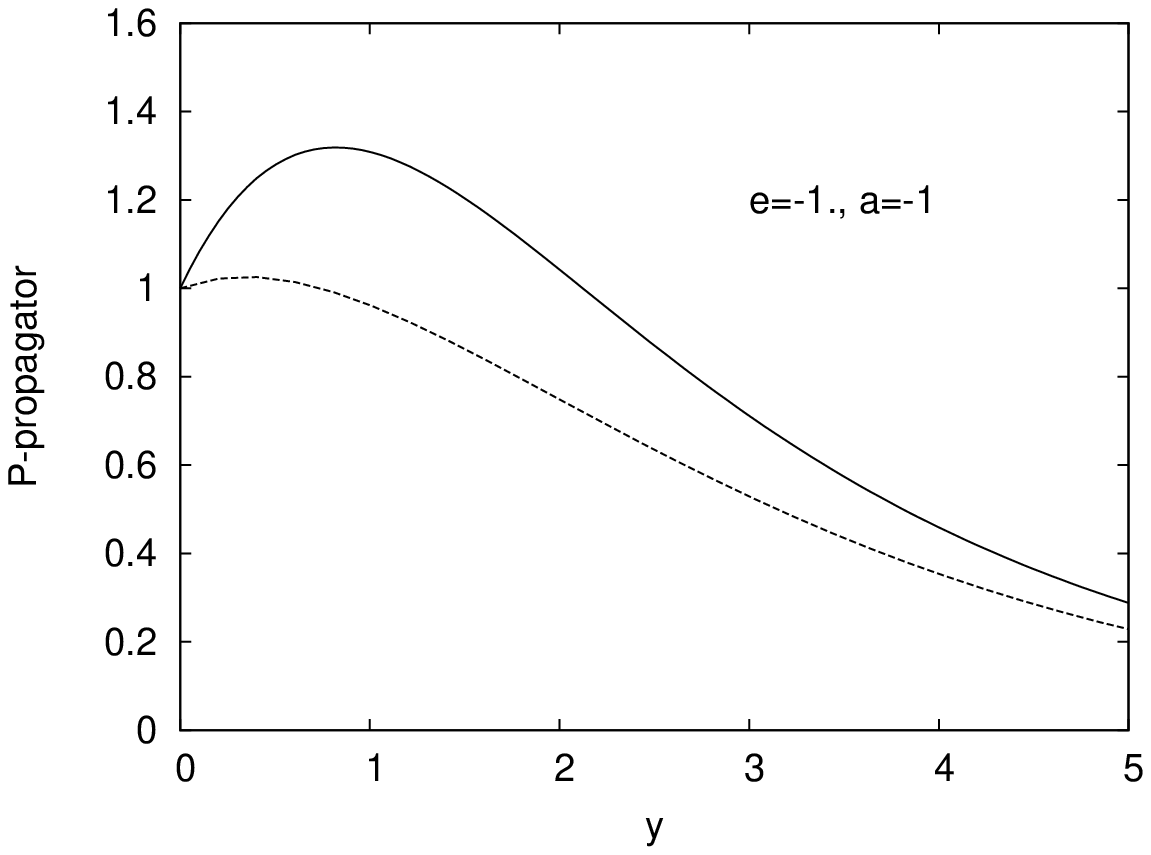}
\includegraphics[width=7.2 cm]{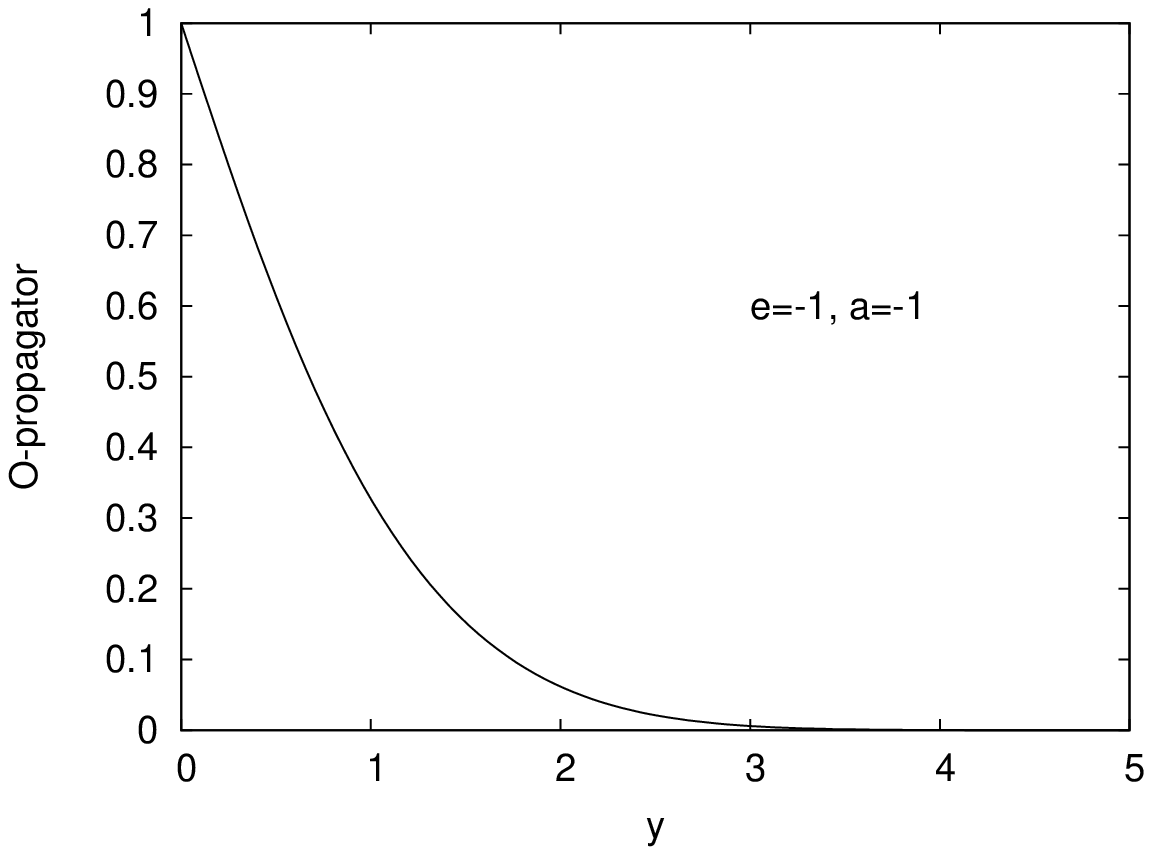}
\caption{The solid curves show pomeron (left panel) and odderon (right panel) propagators
as functions of rapidity for $\mu=1$, $\lambda=1$. The dashed curve in the left panel
shows the pomeron propagator in absence of the odderon (only pomeron loops).}
\label{fig10}
\end{center}
\end{figure}

\begin{figure}[h!p]
\begin{center}
\includegraphics[width=7.2 cm]{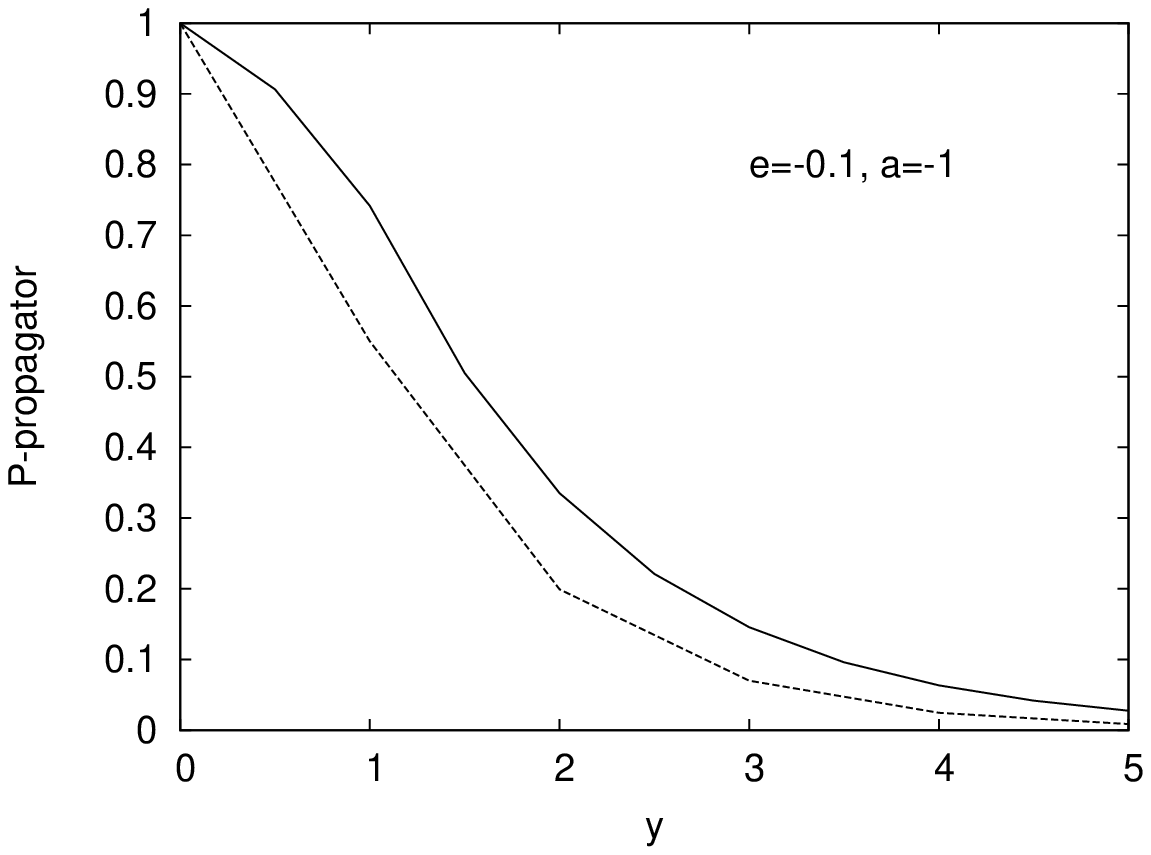}
\includegraphics[width=7.2 cm]{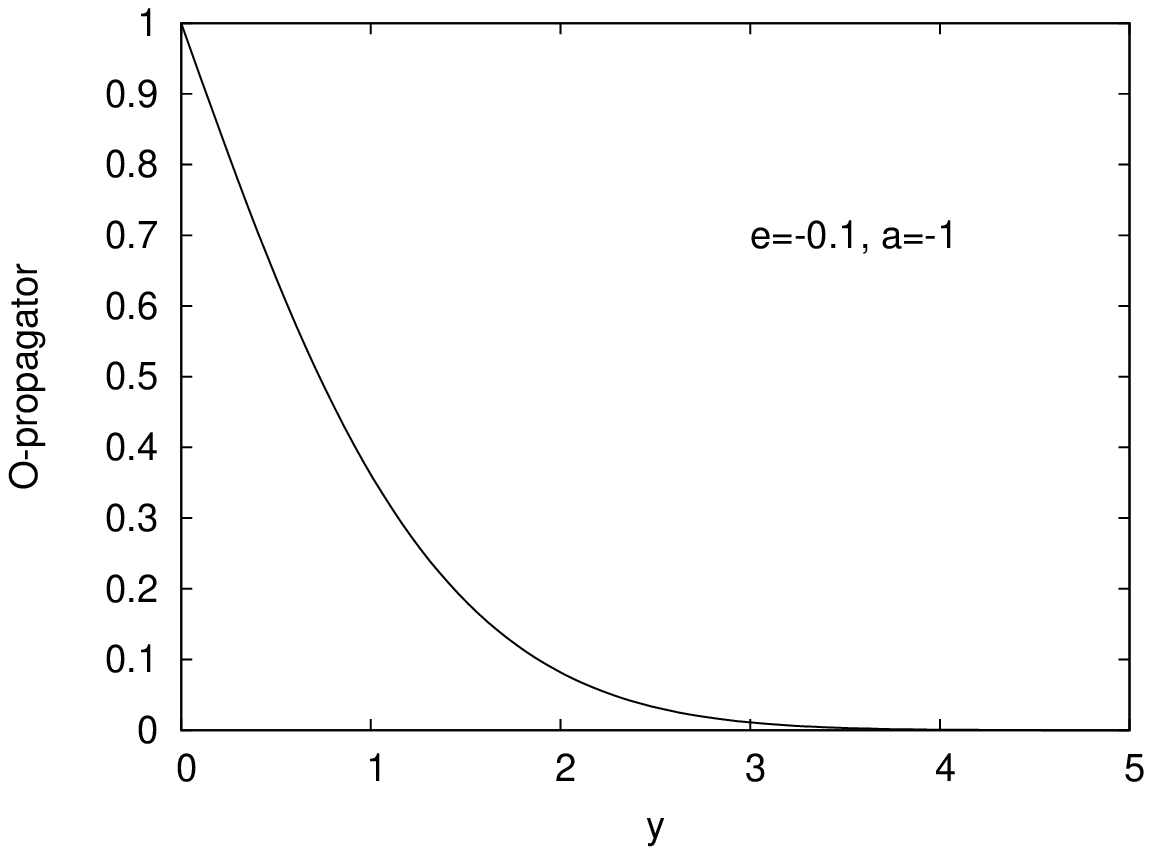}
\caption{The solid curves show pomeron (left panel) and odderon (right panel) propagators
as functions of rapidity $\mu=0.1$, $\lambda=1$. The dashed curve in the left panel
shows the pomeron propagator in absence of the odderon (only pomeron loops).}
\label{fig11}
\end{center}
\end{figure}

At $\lambda\geq 0.04$ evolution by power expansion breaks down.
So the rest plots for the pomeron and odderon propagators shown
in Figs.~\ref{fig7}-\ref{fig11} use evolution by points.
For the pomeron we compared our results including odderon loops
with our old calculations without odderon \cite{braun3,braun1}.
We consider the cases with
$$
(\mu,\lambda)=(1,0.1),\ \ (-1,0.1),\ \ (1,1/3),\ \ (1,1),\ \ (0.1,1),
$$
which fully enough illustrate the dependence on the intercept $1+\mu$
and triple coupling constant $\lambda>0$. The dashed curves in all
plots of the pomeron propagator correspond the results without odderon.


Observing our results we see that the effect of the odderon loops is universally
constructive. As expected they act in the opposite direction compared
to the pomeron loops making the propagator somewhat larger. This effect
is enhanced with the growth of rapidity.
At large values of $\lambda$ the effect of odderon loops starts at quite small
$y$ and becomes very strong, see Figs.~\ref{fig9}-\ref{fig11}.
In all cases at large enough $y$ the pomeron propagator goes to zero.

\newpage
%%5
\section{Quadruple interaction}

\noindent
The conclusion that only the triple interactions are essential
at large energies is based on the renormalization group analysis
\cite{migdal,migdal1}
and is not applicable for the one-dimensional model. To introduce
the quadruple interaction in our theory we shall exploit the idea
that the model can be expressed via the composite fields \Ref{e31}
and add a term to the Hamiltonian
\begin{equation}
H_4=\lambda'\Big(\phi^*\phi^*\phi\phi
+\tilde{\phi}^*\tilde{\phi}^*\tilde{\phi}\tilde{\phi}\Big) .
\label{h4}
\end{equation}
In terms of the pomeron and odderon fields one finds
\begin{equation}
\phi^*\phi^*\phi\phi
=\frac{1}{\sqrt{2}}\Big[({\Phi^*}^2-{\Psi^*}^2)(\Phi^2-\Psi^2)
+4 \Phi^* \Psi^* \Phi\Psi+2i({\Phi^*}^2-{\Psi^*}^2)\Phi\Psi
-2i\Phi^* \Psi^*  (\Phi^2-\Psi^2)\Big] .
\end{equation}
Then the quadruple interaction \Ref{h4} can be written as
\begin{equation}
H_4^{(A)}=\lambda'\Big({\Phi^*}^2\Phi^2 + {\Psi^*}^2\Psi^2
-{\Phi^*}^2\Psi^2-{\Psi^*}^2\Phi^2 + 4\Phi^* \Psi^* \Phi\Psi \Big) .
\end{equation}
Passing to variables $u,w$ \Ref{e53} we have
\begin{equation}
H_4=\lambda'\Big((u^2-w^2)\frac{\pd^2}{\pd u^2}
+(w^2-u^2)\frac{\pd^2}{\pd w^2}+4uw\frac{\pd^2}{\pd u\pd w}\Big) .
\label{h41a}
\end{equation}

One can try to use these form of the interaction to perform the
evolution by points in the space $(u,w)$. Unfortunately the additional
quadruple terms drastically spoil convergence and do not allow to move
above $y=1$.

In contrary, with a non-zero $\lambda'$ evolution by powers allows
to calculate the pomeron and odderon propagators at rather high values
of $\lambda$ and $\lambda'$ at least up to rapidity $y=10$.

In the evolution in powers of $u$ and $w$ this interaction
generates an additional term $f_{nm}^{(4)}$ in \Ref{fnm}.
For $n,m\geq 2$ one finds
\begin{equation}
f^{(4)}_{nm}=\lambda'\Big\{[n(n-1)+m(m-1)+4nm]g_{nm}
-(m+1)(m+2)g_{n-2,m+2}-(n+1)(n+2)g_{n+2,m-2}\Big\}
\label{fnm4}
\end{equation}
and for smaller $n$ or $m$
$$f^{(4)}_{00}=f^{(4)}_{10}=f^{(4)}_{01}=0,\ \ f^{(4)}_{11}=\lambda'g_{11}
$$
$$f^{(4)}_{20}=2\lambda'(g_{20}-g_{02}),\ \ f^{(4)}_{02}=-f^{(4)}_{20}
$$
\begin{equation}
f^{(4)}_{21}=2\lambda'(g_{21}-3g_{03}),
\ \ f^{(4)}_{12}=2\lambda'(g_{12}-3g_{30}) .
\end{equation}

\begin{figure}[ht]
\begin{center}
\includegraphics[width=7.2 cm]{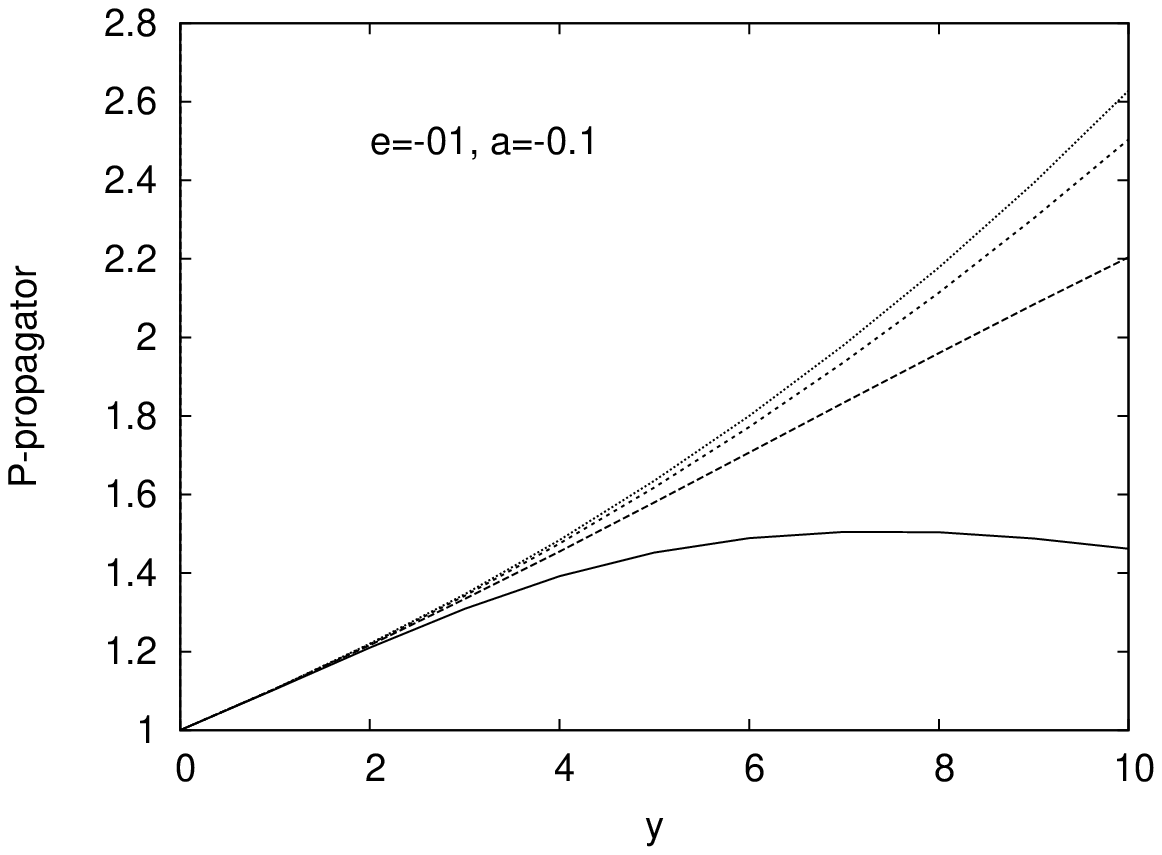}
\includegraphics[width=7.2 cm]{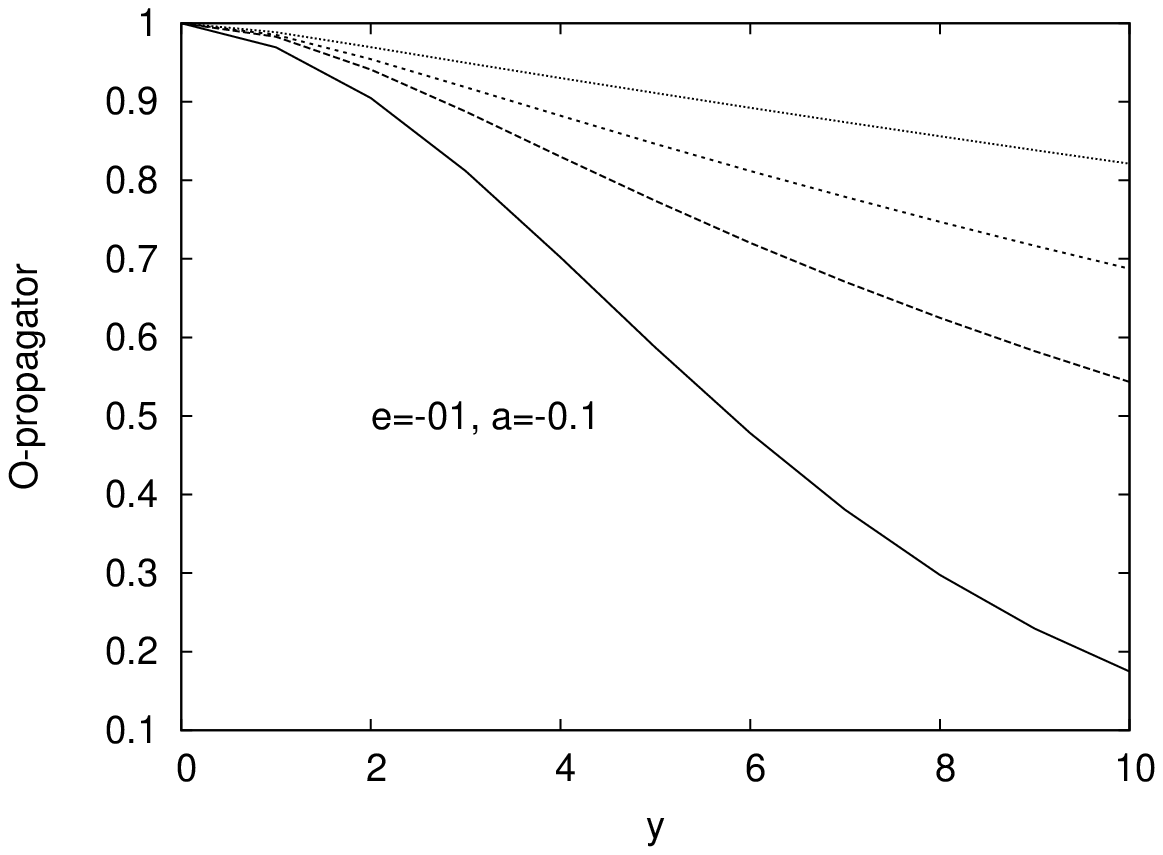}
\caption{Pomeron (left panel) and odderon (right panel) propagators
at different rapidities for $\mu=0.1 $, $\lambda=0.1$ with the quadruple
interaction. Curves from bottom to top correspond to $\lambda'=0$, $0.5$,
$1.0$ and $2.0$.}
\label{quadr}
\end{center}
\end{figure}

In Fig.~\ref{quadr} we show the propagators  at $\mu=0.1$, $\lambda=0.1$
and values of $\lambda'=0$, $0.5$, $1$ and $2$. Convergence becomes better
with the growth of $\lambda'$. In particular at $\lambda'=0$ that is
without the quadruple interaction, the power series method diverges and
the corresponding curves could only be obtained by the point evolution.

Calculation shows that the quadruple interaction somewhat raises
both the pomeron and odderon propagators which are illustrated in
Fig.~\ref{quadr}. The curves show the propagators successively enhanced
as the values of $\lambda'$ are raised: $\lambda'=0, 0.5, 1, 2$.
So the quadruple interaction \Ref{h4} acts in the opposite direction
to the loops coming from the triple interaction providing a constructive
contribution.

%%6
\section{Conclusions}

\noindent
We have proposed an one-dimensional reggeon model for the interaction
of local pomerons and odderons at different energies. The model is a
natural generalization of the well-known ''toy'' model for the pomerons
only and follows the structure of the pomeron-odderon evolution in the
QCD. It allows to study evolution of interacting pomerons and odderons
with quantum effects (loops) fully taken into account and so estimate
these effects and validity of the quasiclassical approximation.

Practical methods to study evolution in rapidity were proposed. In the
general case they can be pursued by the same numerical procedure that
turned out to be successful for the model without odderons, provided
the initial wave function is adequately chosen. In particular, naive
independent initial choice of the pomeron and odderon contribution is
not allowed for evolution, since, probably, it contradicts the asymptotic
properties of the wave function considered as a function of two complex
field describing the pomeron-odderon system.

The found numerical results show that the inclusion of odderon loops
enhances the pomeron propagators but somewhat damps the $pA$ amplitudes.
In particular, the odderon propagator strongly falls with energy
in agreement with conclusions from the QCD. Remarkably the influence
of odderon loops is clearly felt both in the propagators and amplitudes.
also leads to overall damping with energies.

We also studied a certain quadruple interaction choosing it in a form
symmetric in pomerons and odderons. This interaction acts constructively
and raises both the pomeron and odderon propagators.

\section{Acknowledgements}

\noindent
The authors are grateful to the referee for major constructive
comments and suggestions used in the paper.

\end{document}